\documentclass[conference]{IEEEtran}
\IEEEoverridecommandlockouts
\usepackage{cite}
\usepackage{amsmath,amssymb,amsfonts}
\usepackage{algorithmic}
\usepackage{graphicx}
\usepackage{textcomp}
\usepackage{bmpsize}
\usepackage{xcolor}
\usepackage{lipsum}
\usepackage[colorlinks=true,urlcolor=black]{hyperref}
\def\BibTeX{{\rm B\kern-.05em{\sc i\kern-.025em b}\kern-.08em
    T\kern-.1667em\lower.7ex\hbox{E}\kern-.125emX}}

\usepackage{soul}
\usepackage{multirow}
\usepackage{tikz}
\usepackage{epsfig} 
\usepackage{xfrac}
\usepackage{filecontents}
\usepackage{csquotes}
\usepackage{footnote}
\makesavenoteenv{tabular}
\makesavenoteenv{table}
\usepackage{pifont}
\usepackage{pbox}
\usepackage{cancel}
\newtheorem{theorem}{Theorem}

\newtheorem{remark}[theorem]{Remark}

\usepackage{subcaption}
\usepackage{xcolor}
\usepackage{bbm}
\usepackage{graphicx}
\usepackage{soul}
\usepackage{tikz}
\newcommand{\cmark}{\ding{51}} 
\newcommand{\xmark}{\ding{55}} 
\begin{document}

\title{Active Attack Resilience in 5G: A New Take on Authentication and Key Agreement}
\author{
    \IEEEauthorblockN{
        Nazatul H. Sultan\IEEEauthorrefmark{1},
        Xinlong Guan\IEEEauthorrefmark{1},
        Josef Pieprzyk\IEEEauthorrefmark{1}\IEEEauthorrefmark{2},
        Wei Ni\IEEEauthorrefmark{1},
        Sharif Abuadbba\IEEEauthorrefmark{1}, and
        Hajime Suzuki\IEEEauthorrefmark{1}
    }
    \IEEEauthorblockA{\IEEEauthorrefmark{1}CSIRO's Data61, Australia}
    \IEEEauthorblockA{\IEEEauthorrefmark{2}Institute of Computer Science, Polish Academy of Sciences, Poland}
}

\maketitle

\begin{abstract}
As 5G networks continue to expand into critical infrastructure, ensuring secure and efficient user authentication has become more important than ever. The 5G-AKA protocol, standardized by 3GPP in TS 33.501, is the cornerstone of authentication in current 5G deployments. It provides mutual authentication, user privacy, and key secrecy. However, despite its widespread adoption, 5G-AKA suffers from known limitations in both security and performance. While it primarily focuses on protecting privacy against passive attackers, recent studies have highlighted its vulnerabilities to active attacks. Furthermore, it relies on a sequence number-based mechanism to prevent replay attacks, requiring the user device and the core network to remain perfectly synchronized. This stateful design introduces operational complexity, frequent desynchronization issues, and additional communication overhead. More critically, 5G-AKA lacks Perfect Forward Secrecy (PFS), leaving past communications vulnerable if long-term keys are ever compromised - a growing concern in the age of sophisticated adversaries.
\par 
In this paper, we propose an enhanced authentication protocol that builds on the design principles of 5G-AKA while addressing these fundamental shortcomings. First, we present a stateless version of the protocol that eliminates the reliance on sequence numbers, reducing communication complexity while remaining fully compatible with existing SIM cards and network infrastructure. We then extend this design to include PFS with only minimal cryptographic overhead. Both protocols are rigorously analyzed using ProVerif, showing that they meet all major security requirements, including resistance to both passive and active attacks, as well as those outlined by 3GPP and recent academic studies. We also prototype both protocols and evaluate their performance against 5G-AKA and 5G-AKA$'$ (USENIX'21). Our results show that the proposed protocols offer stronger security guarantees with only minor impact on computational costs, making them practical and forward-compatible solutions for 5G and beyond.
\end{abstract}

\begin{IEEEkeywords}
5G; Authentication; Key Agreement; Privacy; Perfect Forward Secrecy
\end{IEEEkeywords}

\section{Introduction}
\label{intro}
With recent advancements in Fifth-Generation (5G) mobile network technology, there has been a significant shift from 3G/4G to 5G. According to Ericsson, by the end of 2029, 5G is expected to cover approximately 85\% of the world's population\footnote{\url{https://www.ericsson.com/en/reports-and-papers/mobility-report/dataforecasts/network-coverage}}. To ensure the security and privacy of 5G users, the Third Generation Partnership Project (3GPP)\footnote{\url{https://www.3gpp.org/}} has proposed various security standards. One such standard is the 5G-Authentication and Key Agreement (5G-AKA) protocol, detailed in 3GPP Technical Specification (TS) 33.501 \cite{3GPP}.

The 5G-AKA protocol involves three main entities: the Subscriber (or User Equipment, UE), the Serving Network (SN), and the Home Network (HN). The subscriber refers to the mobile user connected to the 5G network via a Universal Integrated Circuit Card (UICC), commonly known as a SIM card.\footnote{We use “subscriber” and “UE” interchangeably throughout the paper.} The SN and HN represent the base station of a network carrier and the subscriber’s home carrier, respectively. 5G-AKA provides essential security features, including mutual authentication and the establishment of a secure session key between the UE and the network, thereby ensuring secure communication.

The 5G-AKA protocol builds on the earlier 3G/4G-AKA protocols, offering improved privacy through the use of public-key encryption to protect the subscriber's permanent identity (i.e., Subscription Permanent Identifier, SUPI) \cite{Basin2018}. As a foundational component of 5G security, 5G-AKA has undergone extensive security evaluations \cite{Basin2018, Cremers2019, Miller2022}, similar to its predecessors \cite{Arapinis2012, Khan2015}. These analyses have uncovered the following major inherent issues in the 5G-AKA protocol (we also discuss it in detail in Section \ref{sec:5g-aka-analysis}):

\paragraph{Lack of Active Attack Resistance} The 5G-AKA protocol is primarily designed to defend against passive attackers, who only eavesdrop on communications. However, recent studies \cite{Basin2018, Borgaonkar2019, Koutsos2019} have shown that it remains vulnerable to active attackers, who can manipulate, replay, or inject messages into the protocol flow, thereby compromising user privacy. Prior research \cite{Arapinis2012, Fouque2016, Borgaonkar2019, Koutsos2019} have also demonstrated privacy attacks—especially against subscriber unlinkability—that allow adversaries to distinguish and track specific users. These threats are particularly concerning for high-profile individuals (e.g., journalists, activists, or political figures). With the emergence of open-source 5G platforms such as \cite{Free5G, Open5gcore, Openairinterface}, the feasibility of active attacks has become even more realistic \cite{Wang2021}.

\paragraph{Lack of Perfect Forward Secrecy (PFS)} The 5G-AKA does not offer PFS \cite{Basin2018}. If the UE’s long-term secret key is compromised, an adversary can compute past session keys and decrypt previously captured traffic. This is particularly a serious concern in an era of advanced persistent threats where long-term key compromise is increasingly plausible.

\paragraph{Inefficient Replay Attack Prevention} The 5G-AKA protocol uses a sequence number-based mechanism to prevent replay attacks. This requires synchronization of the sequence number between the UE and HN, making the protocol stateful. This stateful design introduces additional challenges such as operational complexity, frequent desynchronization issues, and communication overhead for resynchronization \cite{Basin2018, 3GPP}.
\par 
Furthermore, the comprehensive formal analysis by Basin \emph{et al.} \cite{Basin2018} (CCS'18), based on 3GPP’s security specifications, also highlights several underspecified requirements and inherited weaknesses from 3G/4G-AKA, such as the lack of key confirmation. The authors recommend designing a protocol specifically tailored for 5G, rather than extending legacy systems, due to outdated assumptions and constraints (e.g., lack of robust pseudo-random number generators) that have since been addressed in modern 5G infrastructure.
\par 
To address these challenges, several studies—including \cite{Koutsos2019, Wang2021, Braeken2020, Munilla2021}—have proposed enhancements to the 5G-AKA protocol. However, most of these solutions fall short of delivering all essential security properties within a unified framework. In particular, they often fail to simultaneously defend against active attacks, provide PFS, and implement a robust replay protection mechanism that does not rely on sequence numbers—all while maintaining compatibility with existing 5G SIM cards. We discuss these works in more detail in the Related Work section.

\paragraph*{Our Contribution} 
In this paper, we delve deeper into the issues present in the 5G-AKA protocol and propose an enhanced and secure version of the AKA protocol tailored for 5G. We summarize our key contributions below. 
\begin{itemize}
    \item We design an efficient and secure AKA protocol for 5G that fulfills all security guarantees outlined in 3GPP TS 33.501 \cite{3GPP}. Our protocol, referred to as Protocol I, also addresses the additional underspecified security requirements discussed in \cite{Basin2018}. Importantly, our protocol is resilient against both passive and active attackers. While following a message flow pattern similar to 5G-AKA, we modify the challenge-response method to prevent replay attacks and ensure mutual authentication between the subscriber and the HN. This new method avoids the inefficient sequence number-based replay attack prevention techniques used in 5G-AKA while maintaining compatibility with the existing 5G infrastructure, especially 5G SIM cards.
    
    \item We propose an extension of our protocol I to accommodate the additional property of PFS with the expense of a few additional lightweight cryptographic operations. In our extension, referred to as Protocol II, we efficiently introduce an ephemeral Diffie-Hellman (DH) key exchange method along with a modified challenge-response process from Protocol I. This extension supports all essential security requirements of our protocol I, including resistance to both passive and active attacks.
    \item We perform a comprehensive formal security analysis using the state-of-the-art symbolic model-based automated security verification tool \emph{ProVerif} \cite{Blanchet2014}. Our formal analysis indicates both of our protocols (Protocol I and Protocol II) support all essential security requirements outlined in 3GPP TS 33.501 \cite{3GPP} and \cite{Basin2018}, including mutual authentication, secrecy, and active attack resistance. Additionally, our analysis shows that the extension protocol II offers PFS. All code is available at \url{https://anonymous.4open.science/r/AKA-5G-ProVerif-9378/}.
    
    \item We provide a comprehensive comparison of our protocols with the 5G-AKA protocol \cite{3GPP} and 5G-AKA$'$ protocol \cite{Wang2021} (USENIX'21). Additionally, we conduct experiments using the Crypto++ library \cite{crypto++} to validate our findings. All code is available at \url{https://anonymous.4open.science/r/AKA-5G-E8B4/}.
  
\end{itemize}

\begin{table}[!t]
\centering
\caption{Functionality and Security Comparison between Our Protocols and Other Notable Works } \label{table:functionality-comparison}
\tiny
\begin{tabular}{|l|c|c|c|c|cl|}
\hline
\multirow{2}{*}{} 
& \multirow{2}{*}{\begin{tabular}[c]{@{}c@{}}Mutual\\Authentication\end{tabular}} 
& \multirow{2}{*}{\begin{tabular}[c]{@{}c@{}}Active Attack\\Resistance\end{tabular}} 
& \multirow{2}{*}{PFS} 
& \multirow{2}{*}{\begin{tabular}[c]{@{}c@{}}SQN\\Resync\end{tabular}} 
& \multicolumn{2}{c|}{Compatibility$^{\#}$} \\ \cline{6-7} 
& & & & 
& \multicolumn{1}{c|}{USIM} 
& \multicolumn{1}{c|}{SN} \\ \hline
  DDAIP \cite{Khan2018}  & \multicolumn{1}{l|}{\cmark} & \multicolumn{1}{l|}{\xmark} & \multicolumn{1}{l|}{\xmark} & \multicolumn{1}{l|}{\checkmark} & \multicolumn{1}{l|}{\cmark} & \cmark \\ \hline
MultiIMSI-4G \cite{Khan2015}  & \multicolumn{1}{l|}{\cmark} & \multicolumn{1}{l|}{\xmark} & \multicolumn{1}{l|}{\xmark} & \multicolumn{1}{l|}{\checkmark} & \multicolumn{1}{l|}{\cmark} &  \cmark\\ \hline
DefeatCatchers\cite{van2015}  & \multicolumn{1}{l|}{\cmark} & \multicolumn{1}{l|}{\xmark} & \multicolumn{1}{l|}{\xmark} & \multicolumn{1}{l|}{\checkmark} & \multicolumn{1}{l|}{\cmark} & \cmark \\ \hline
PrivacyThreats-5G\cite{Borgaonkar2019}  & \multicolumn{1}{l|}{\cmark} & \multicolumn{1}{l|}{\xmark} & \multicolumn{1}{l|}{\xmark} & \multicolumn{1}{l|}{\checkmark} & \multicolumn{1}{l|}{\cmark} &  \cmark\\ \hline
 AKA-FS \cite{Arkko2015}$^*$  & \multicolumn{1}{l|}{\cmark} & \multicolumn{1}{l|}{\cmark} & \multicolumn{1}{l|}{\cmark} & \multicolumn{1}{l|}{\checkmark} & \multicolumn{1}{l|}{\xmark} &  \xmark\\ \hline
 TSA-5G \cite{Liu2018}$^*$  & \multicolumn{1}{l|}{\cmark} & \multicolumn{1}{l|}{\cmark} & \multicolumn{1}{l|}{\cmark} & \multicolumn{1}{l|}{\checkmark} & \multicolumn{1}{l|}{\xmark} & \xmark \\ \hline
  5G-AKA-FS \cite{You2024}$^*$  & \multicolumn{1}{l|}{\cmark} & \multicolumn{1}{l|}{\cmark} & \multicolumn{1}{l|}{\cmark} & \multicolumn{1}{l|}{\checkmark} & \multicolumn{1}{l|}{\xmark} & \xmark \\ \hline
  Symmetric-AKA \cite{Braeken2020}  & \multicolumn{1}{l|}{\cmark} & \multicolumn{1}{l|}{\xmark} & \multicolumn{1}{l|}{\cmark} & \multicolumn{1}{l|}{--} & \multicolumn{1}{l|}{\cmark} &  \cmark\\ \hline
   Beyond-5G \cite{Damir2022}  & \multicolumn{1}{l|}{\cmark} & \multicolumn{1}{l|}{\cmark} & \multicolumn{1}{l|}{\cmark} & \multicolumn{1}{l|}{--} & \multicolumn{1}{l|}{\xmark} & \xmark \\ \hline
AKA$^+$ \cite{Koutsos2019} & \multicolumn{1}{l|}{\cmark} & \multicolumn{1}{l|}{\cmark} & \multicolumn{1}{l|}{\xmark} & \multicolumn{1}{l|}{\checkmark} & \multicolumn{1}{l|}{\xmark} & \xmark \\ \hline
5G-AKA$'$ \cite{Wang2021} & \multicolumn{1}{l|}{\cmark} & \multicolumn{1}{l|}{\cmark} & \multicolumn{1}{l|}{\xmark} & \multicolumn{1}{l|}{\checkmark} & \multicolumn{1}{l|}{\cmark} & \cmark \\ \hline
5G-AKA \cite{3GPP} & \multicolumn{1}{l|}{\cmark} & \multicolumn{1}{l|}{\xmark} & \multicolumn{1}{l|}{\xmark} & \multicolumn{1}{l|}{\checkmark} & \multicolumn{1}{l|}{\cmark} & \cmark \\ \hline
Our Protocol I  & \multicolumn{1}{l|}{\cmark} & \multicolumn{1}{l|}{\cmark} & \multicolumn{1}{l|}{\xmark} & \multicolumn{1}{l|}{--} & \multicolumn{1}{l|}{\cmark} & \cmark \\ \hline
Our Protocol II & \multicolumn{1}{l|}{\cmark} & \multicolumn{1}{l|}{\cmark} & \multicolumn{1}{l|}{\cmark} & \multicolumn{1}{l|}{--} & \multicolumn{1}{l|}{\cmark} &  \cmark\\ \hline
\end{tabular}
\\ 
\scriptsize{\cmark indicates that the property is supported by the protocol; \xmark indicates that the property is not supported by the protocol; \checkmark indicates the functionality is required by the protocol; -- indicates the functionality is not required by the protocol; PFS means Perfect Forward Secrecy; SQN Resync means Sequence number re-synchronization process; $^*$\cite{Arkko2015}, \cite{Liu2018}, and \cite{You2024} have not provided thorough formal security analyses; $\#$: these properties have not yet been experimentally verified.}
\end{table}

\section{Related Work}
\label{sec:related_work}
The 5G-AKA protocol has undergone thorough scrutiny, revealing several limitations, particularly regarding subscribers' privacy. Several existing studies have identified these issues and proposed various mitigation techniques. In this section, we briefly introduce existing efforts aimed at enhancing the security and privacy of the 5G-AKA protocol, including those that have conducted formal security analyses. Table \ref{table:functionality-comparison} presents a summarized comparison of the security and functionality between our protocols and existing notable works.

\par 
In \cite{Khan2018}, a pseudonym mechanism was proposed to mitigate linkability (or distinguishability) attacks on subscriber privacy in 5G. This mechanism allows subscribers to utilize multiple identities instead of a single permanent identity (i.e., SUPI). A similar pseudonym-based mechanism was also proposed for 3G/4G-AKA protocols in \cite{van2015, Khan2015}. While this technique enhances subscriber privacy to some extent, it does not offer protection against active attackers \cite{Wang2021}. In \cite{Borgaonkar2019}, an attack was presented that could disclose the sequence numbers of targeted subscribers. This attack exploits weaknesses in the sequence number protection mechanism in the 5G-AKA. Although three countermeasures were proposed, unfortunately, none of them can prevent the encrypted SUPI replay attack demonstrated in \cite{Fouque2016}, \cite{Koutsos2019}. 
\par 
In \cite{Koutsos2019}, the AKA$^+$ scheme was proposed for 5G, aiming to withstand all known types of attacks on subscribers' privacy. It implements two key strategies to mitigate privacy breaches in the 5G-AKA protocol. Firstly, it delays the re-synchronization message, which is utilized when the subscriber and the HN become unsynchronized regarding the common sequence number. Secondly, it introduces a challenge message to the subscriber before initiating authentication. However, AKA$^+$ brings about several significant modifications to the 5G-AKA protocol. Most notably, it renders the existing USIM (Universal Subscriber Identity Module) commands incompatible with AKA$^+$, necessitating the replacement of USIM cards with new ones \cite{Wang2021}.

\par 
In \cite{Wang2021}, the 5G-AKA$'$ was proposed with the aim of mitigating all known privacy-related attacks in 5G-AKA while making minimal modifications. The key idea is to verify the freshness of the challenge message from the HN before the subscriber processes other operations. Unlike AKA$^+$, 5G-AKA$'$ is compatible with existing USIM cards. However, 5G-AKA$'$ inherits other limitations of the 5G-AKA protocol, including sequence number resynchronization and the lack of forward secrecy if the long-term secret key of the subscriber and the private key of the HN are compromised. In \cite{Arkko2015}, \cite{Liu2018}, and \cite{You2024}, three DH key exchange-based schemes have been proposed. However, similar to AKA$^+$, both \cite{Arkko2015} and \cite{Liu2018} are also incompatible with existing USIM cards and SNs' implementations. Further, the works \cite{Arkko2015}, \cite{Liu2018}, and \cite{You2024} have not provided thorough security analyses. This means that the reader cannot be sure whether or not the claimed security goals are indeed achieved.

\par 
In \cite{Braeken2020}, a symmetric key-based AKA protocol is proposed for 5G. The protocol does not use any public-key cryptographic primitives and supports some of the essential security requirements such as anonymity, unlinkability, mutual authentication, and confidentiality. However, the work \cite{Braeken2020} is vulnerable to location confidentiality attacks, as shown in \cite{Munilla2021}. In \cite{Damir2022}, a quantum-safe AKA protocol is proposed for 5G. The scheme uses post-quantum KEM (Key Encapsulation Mechanism) instead of relying on the public-key cryptographic primitives. It is incompatible with existing USIM cards and SNs' implementations, similar to \cite{Koutsos2019}, \cite{Arkko2015}, and \cite{Liu2018}.

\par 
Apart from proposing new schemes to enhance 5G-AKA protocols, extensive security analyses have been conducted on the existing 5G-AKA protocol. The authors in \cite{Basin2018} provide a detailed formal security analysis of 5G-AKA, offering a comprehensive definition of its security and privacy properties based on 3GPP's specifications. Utilizing Tamarin Prover \cite{Tamarin}, the analysis identifies several underspecified security requirements within the 5G-AKA protocol, including the lack of forward secrecy, vulnerability to active attackers, and absence of key confirmation. Similarly, the study in \cite{Cremers2019} presents another thorough formal analysis, considering all four essential components of a real-world 5G-AKA protocol and various channel-compromised scenarios. This analysis, also conducted using Tamarin Prover \cite{Tamarin}, reveals that the security of the 5G-AKA protocol relies on underspecified assumptions regarding the inner workings of underlying channels, potentially leading to security-critical race conditions. Furthermore, a recent formal analysis of 5G-AKA is outlined in \cite{Miller2022}, focusing specifically on the different phases within the authenticated key agreement procedure and their impact on critical mobile-network objects such as Protocol Data Unit (PDU) sessions. In addition to the formal analysis of the 5G-AKA protocol, its predecessors, such as 3G/4G, have also undergone extensive scrutiny in \cite{Arapinis2012}, \cite{Borgaonkar2019}, and \cite{3G_formal_Analy}.

\begin{figure*}[!t]
    \centering
	\fbox{\scalebox{7}{\includegraphics[width=2.1cm, height=2.1cm]{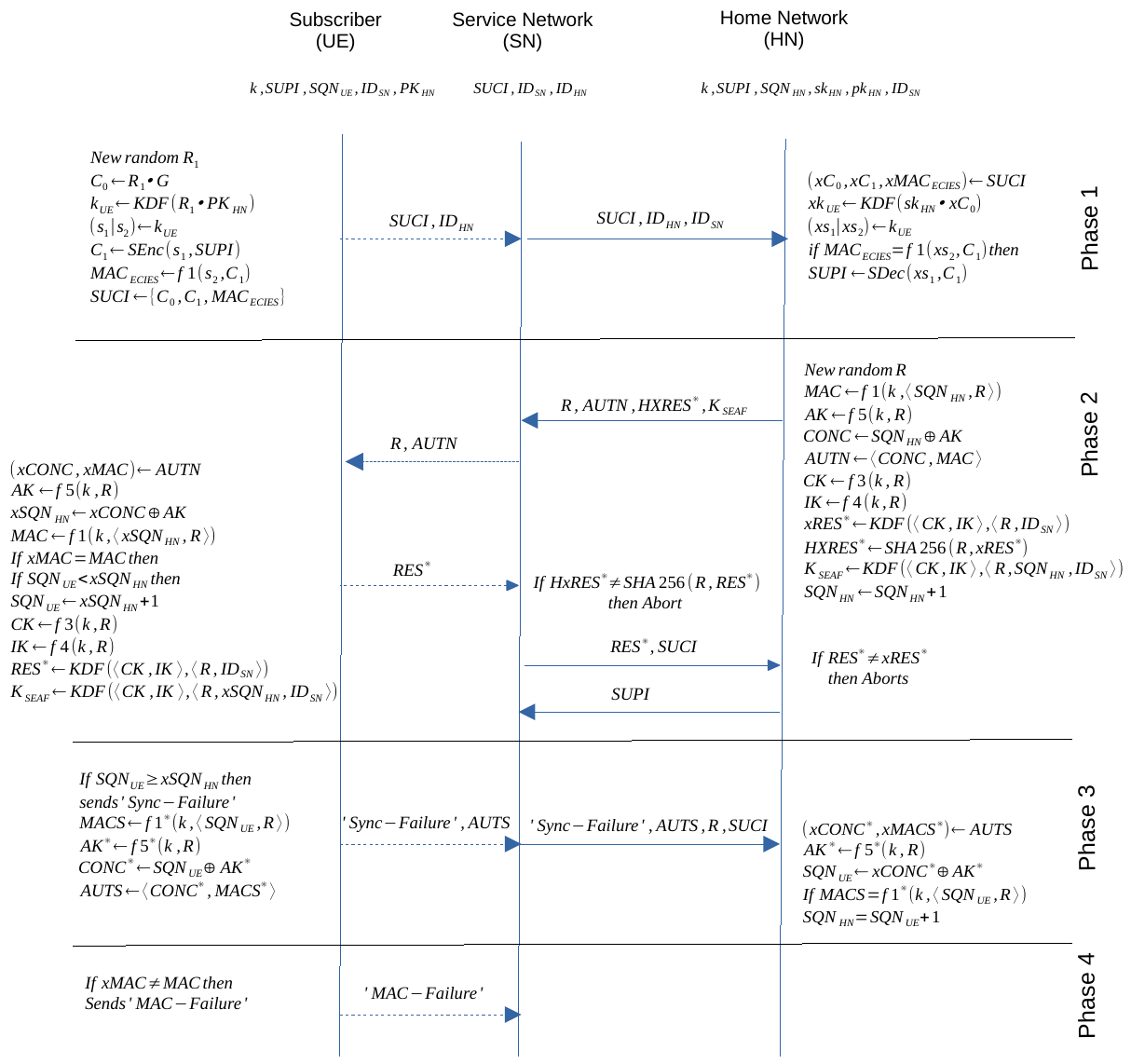}}}
	\caption{A high-level overview of the 5G-AKA protocol, where dotted and solid arrows represent open channels and authenticated secure channels, respectively.}		
    \label{fig:5G-AKA}
 \end{figure*} 
\section{5G-AKA Protocol}
\label{sec:5G-AKA}
In this section, we present an overview of the 5G-AKA protocol as defined by the 3GPP TS 33.501 standard \cite{3GPP}, followed by a brief discussion of its associated privacy issues in Section \ref{sec:5g-aka-privacy-issue}. This overview aims to highlight the distinctions between 5G-AKA and our proposed protocols, as detailed in the subsequent sections of this paper. Additionally, it emphasizes the need for a more secure AKA protocol within the 5G framework. Following the examples of \cite{Basin2018} and \cite{Wang2021}, we did not consider all components of the HN for simplicity.
\par 
The 5G-AKA protocol uses several cryptographic primitives such as hash functions $f1, f2, f3, f4, f5, f1^*, f5^*$, the Secure Hash Algorithm-256 ($\mathrm{SHA256}$), the Elliptic Curve Integrated Encryption Scheme (ECIES) for key encapsulation (please refer to Appendix \ref{sec:ECIES} for more details), and the Key Derivation Function ($\mathrm{KDF}$). More details on the cryptographic primitives can be found in 3GPP TS 33.501 \cite{3GPP}.
\par 
As mentioned earlier, the 5G-AKA protocol involves three primary entities: UE, SN, and HN. The subscriber or UE represents the mobile device user with the USIM card and connects to the 5G network to access its services. The USIM card typically contains essential information, including the Subscription Permanent Identifier (\textsc{SUPI}), the long-term secret key ($\mathsf{k}$), the sequence number ($\mathsf{SQN_{UE}}$), and the HN's public key ($\mathsf{PK_{HN}}$). The SN is the network carrier that the subscriber connects to, while the HN is the subscriber's own network carrier. The main functionality of the 5G-AKA protocol is to provide mutual authentication and establish a session key between the entities. A high-level overview of the 5G-AKA protocol is shown in Figure \ref{fig:5G-AKA}. We provide a detailed overview of the 5G-AKA protocol in Appendix \ref{appendix:5G-AKA}. Here, we briefly touch upon each phase. 5G-AKA consists of four phases: \emph{Initiation} (Phase 1), \emph{Challenge-Response} (Phase 2), \emph{Sequence Number Re-synchronization} (Phase 3), and \emph{MAC Failure} (Phase 4).

\paragraph*{Initiation (Phase 1)} In the Initiation phase, the subscriber (i.e., the UE) sends the encrypted SUPI using ECIES, which is represented as SUCI (Subscription Concealed Identifier), to the HN via the radio channel through an SN.
\paragraph*{Challenge-Response (Phase 2)} In the Challenge-Response phase, the HN chooses a random value $R$. It computes $\mathrm{AUTN}$, which consists of the $\mathsf{MAC}$ (Message Authentication Code) for the $R$ and the XOR ($\oplus$) of the sequence number $\mathsf{SQN_{HN}}$ for privacy. The subscriber performs two tasks upon receiving the challenge message from the HN. First, it checks the $\mathsf{MAC}$ to authenticate and verify the integrity of the challenge $R$. Second, the $\mathsf{SQN_{HN}}$ is used to check the freshness of the challenge $R$ from the HN to prevent replay attacks. The subscriber checks the freshness of the challenge message by comparing the received $\mathsf{SQN_{HN}}$ with its own sequence number $\mathsf{SQN_{UE}}$. Please note that, as per the 3GPP TS 33.50 \cite{3GPP}, $\mathsf{SQN_{HN}}$ and $\mathsf{SQN_{UE}}$ should remain synchronized at all times.
\paragraph*{Sequence Number Re-synchronization (Phase 3)} If this comparison of the sequence numbers fails, i.e., if $\mathsf{SQN_{HN}}$ and $\mathsf{SQN_{UE}}$ are unsynchronized, the subscriber sends a \textsc{Sync\_Failure} message, along with $\mathrm{AUTN}$, to the HN for re-synchronization of its sequence number $\mathsf{SQN_{UE}}$ with the HN’s $\mathsf{SQN_{HN}}$. Afterward, the subscriber returns to the Initiation phase to restart the process from the beginning.

\paragraph*{MAC Failure (Phase 4)} If the $\mathsf{MAC}$ check fails, the subscriber outputs a \textsc{MAC\_Failure} message and returns to the Initiation phase to restart the process.
If both the $\mathsf{MAC}$ and $\mathsf{SQN_{HN}}$ checks are successfully completed, the subscriber generates a response message $\mathsf{RES}$ for the challenge $R$, computes the key material to generate the anchor keys (i.e.,  $\mathsf{K_{SEAF}}$), and sends $\mathsf{RES}$ to the HN. 

\subsection{Analysis of 5G-AKA Protocol Limitations}\label{sec:5g-aka-analysis}
In this section, we briefly revisit the key challenges of the 5G-AKA protocol introduced in Section \ref{intro}.

\subsubsection{Privacy Threats in 5G-AKA from Active Attackers}\label{sec:5g-aka-privacy-issue}
The 5G-AKA protocol is primarily vulnerable to three types of privacy threats that compromise subscriber privacy. These threats aim to violate the property of unlinkability, which safeguards genuine subscribers from being uniquely identified or distinguished. Below, we summarize the three linkability attacks targeting subscribers, with detailed explanations provided in Appendix \ref{appendix:5G-AKA-Privacy-Issue}.
\paragraph*{Failure Message Linkability Attack \cite{Arapinis2012}} This attack aims to distinguish a targeted subscriber from others by replaying records of $\left<R, \mathrm{AUTN}\right>$ to all subscribers in the vicinity and analyzing their response patterns.   
    \paragraph*{Sequence Number Inference Attack \cite{Borgaonkar2019}} The objective of this attack is to infer information about the targeted subscriber's sequence number $\mathrm{SQN_{UE}}$ by repeatedly replaying previously captured $\left<R, \mathrm{AUTN}\right>$ tuples.   
    \paragraph*{Encrypted SUPI Replay Attack \cite{Fouque2016, Koutsos2019}} This attack seeks to identify the targeted subscriber by replaying a captured $\mathrm{SUCI}$ during the Initiation phase across all subscriber sessions, then analyzing the responses to the corresponding challenge messages from the Home Network (HN).

\subsubsection{Lack of PFS} As shown in Figure \ref{fig:5G-AKA}, the anchor key $\mathsf{K_{SEAF}}$ is derived from the cipher key $\mathsf{CK}$, integrity key $\mathsf{IK}$, the random challenge $R$, the sequence number $\mathsf{SQN_{HN}}$, and the serving network identity $\mathsf{ID_{SN}}$. The keys $\mathsf{CK}$ and $\mathsf{IK}$ are generated by the home network based on the subscriber’s long-term secret key $\mathsf{k}$ and the random number $R$, where $R$ is sent over the open channel to the subscriber. Additionally, the sequence number $\mathsf{SQN_{HN}}$ can be recovered from the concealed value $\mathsf{CONC}$ by first deriving the anonymity key $\mathsf{AK}$ using the long-term key $\mathsf{k}$ and $R$. Since both $\mathsf{R}$ and $\mathsf{CONC}$ are transmitted in the clear, an attacker with access to the long-term key $\mathsf{k}$ can recover $\mathsf{SQN_{HN}}$. This implies that if the long-term key $\mathsf{k}$ is ever compromised and the attacker has recorded previous protocol runs, they can retroactively compute $\mathsf{CK}$, $\mathsf{IK}$, and eventually $\mathsf{K_{SEAF}}$. As a result, both past and future session keys are at risk, meaning the 5G-AKA protocol does not offer PFS.

\subsubsection{Inefficient SQN-based Replay Attack Prevention} As shown in Phase 2 of Figure~\ref{fig:5G-AKA}, the subscriber compares its stored sequence number $\mathsf{SQN_{UE}}$ with the sequence number $\mathsf{xSQN_{HN}}$ received from the HN. For authentication to succeed, these two values must match. Both the subscriber and the HN increment their respective sequence numbers during the authentication process. However, if the sequence numbers become misaligned- due to network issues, message loss, or delays- the authentication will fail. This failure triggers Phase 3, the sequence number (SQN) resynchronization process, which aims to restore synchronization between the two parties. This additional step increases communication overhead and adds complexity to the protocol. Furthermore, as discussed in Section~\ref{sec:5g-aka-privacy-issue}, the use of sequence numbers can introduce privacy risks. In particular, it opens the door to attacks such as the Sequence Number Inference Attack \cite{Borgaonkar2019}, where an attacker may deduce information about a subscriber's activity based on sequence number behavior.


\begin{figure}[!t]
    \centering
	\fbox{\scalebox{6}{\includegraphics[width=1cm, height=.4cm]{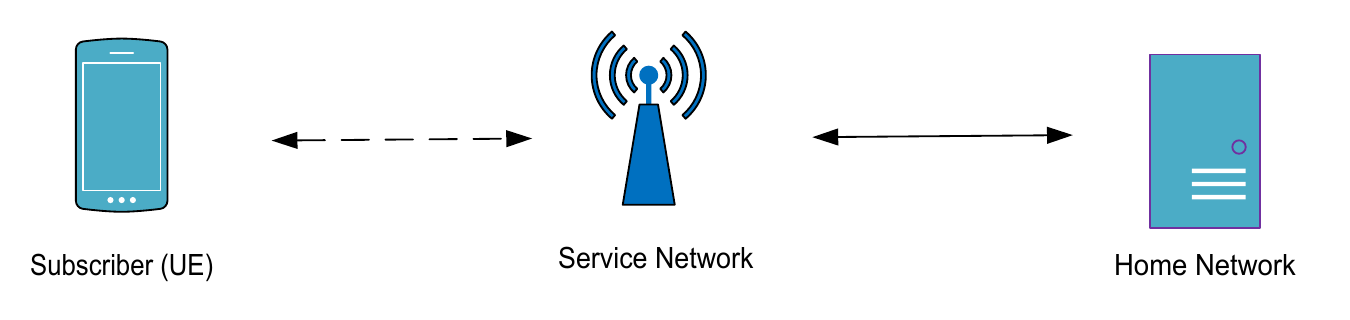}}}
	\caption{Overall Architecture, where dotted and solid arrows represent open channels and authenticated secure channels, respectively.}		
    \label{fig:architecture}
 \end{figure}

\section{System Model, Threat Model \& Security Requirements}
\label{sec:system_threat_model_security_requirements}
In this section, we present the system model, threat model, and security requirements for our protocols, which are primarily based on the 3GPP standards (3GPP TS 33.501 \cite{3GPP}).
\subsection{System Model}
\label{sec:system_model}
Our protocols consider three broad roles in the system model similar to Basin \emph{et al.} \cite{Basin2018}: \emph{Subscriber} or UE, HN, and SN. Figure \ref{fig:architecture} shows the entities/roles and channels involved in our protocols. The subscribers (UEs) represent smartphones or IoT devices equipped with USIMs. The USIM is a cryptographic chip that stores subscriber-related information, such as the long-term secret key ($\mathsf{k}$) and the unique identity of the subscriber, known as ``Subscription Permanent Identifier" or $\mathsf{SUPI}$. Each subscriber is registered with an HN. The HN is typically the network carrier of the subscriber, which maintains the database of the subscribers and also performs authentication before providing access to its services. It also stores the same subscriber-related information as the USIM for each registered subscriber. In the real world, the HN consists of multiple sub-entities, such as the Authentication Server Function (AUSF), which authenticates subscribers, and Unified Data Management (UDM), which computes keying material and authentication-related data for the AUSF. The UDM consists of the Authentication Credential Repository and Processing Function (ARPF) and Subscription Identifier De-concealing Function (SIDF). The primary purpose of ARPF is to store subscribers' secret credentials and compute authentication-related cryptographic parameters. On the other hand, SIDF contains the HN's private key, which recovers the plaintext $\mathrm{SUPI}$ of subscribers from its encrypted version $\mathrm{SUCI}$. For simplicity and without compromising overall security, we consider AUSF and UDM as a single entity, which is HN. The SNs are the actual network carriers to which subscribers connect and get access to the cellular network. When subscribers are within their HN coverage area, the HN also serves as the SN. In roaming scenarios, however, the SN is operated by a different network provider. We note that, similar to the 5G-AKA protocol, our proposed protocols are applicable in both roaming and non-roaming scenarios.

\begin{remark}
    In our model, we abstract the UE and USIM as a single logical entity, consistent with previous works (e.g.,~\cite{Basin2018, Wang2021}) and common in formal protocol analysis. However, we acknowledge that in practical deployments, SUCI (i.e., the encrypted SUPI) may be generated either on the USIM or on the host UE, depending on the SIM profile and configuration, as specified in 3GPP TS~31.121~\cite{3gpp-ts-31.121} and TS~33.501~\cite{3GPP}. When SUCI is generated on the UE, the SUPI must traverse the SIM--UE interface in plaintext, which introduces a potential attack surface. Recent works such as SecureSIM~\cite{Zhao2021} and SIMurai~\cite{Lisowski2024} have demonstrated that this interface can be vulnerable to runtime abuse, man-in-the-middle attacks, and software-level compromise. 
    \par 
    Our analysis assumes that SUCI is generated within the USIM, which is the recommended and more secure configuration according to 3GPP specifications. We highlight this assumption to clarify the trust boundary of the subscriber entity in Protocols~I and~II, presented in the following section. In cases where SUCI is instead computed on the UE, we assume the presence of existing protection mechanisms, such as multi-factor access control and runtime isolation, as proposed in SecureSIM~\cite{Zhao2021} and SIMurai~\cite{Lisowski2024}.

\end{remark}



\subsection{Threat Model \& Security Assumptions}
\label{subsec:threat_model}
Our threat model incorporates all the requirements outlined in the 3GPP standards (TS 33.501 \cite{3GPP}). Additionally, our protocols take into account the additional security requirements highlighted in \cite{Basin2018} and \cite{Cremers2019}. 

\paragraph*{Assumption on Channels} 
In accordance with 3GPP standards, the communication channel between SNs and HNs is considered authenticated and secure (i.e., a private channel). This ensures that any attempt by an attacker to eavesdrop, insert, or modify messages within the channel will be prevented and detected. However, the communication channel between the subscriber and the SN is considered insecure or open, allowing passive attackers to eavesdrop and active attackers to manipulate, intercept, and inject messages. Research by Basin et al. \cite{Basin2018} and Cremers et al. \cite{Cremers2019} emphasizes the necessity for a binding channel between SNs and HNs, where each message is associated with a unique session ID to maintain security. Our protocols operate under the assumption of such channel binding between SNs and HNs.

\paragraph*{Assumption on Cryptographic Functions} Our protocols do not require all the cryptographic primitives outlined in 3GPP TS 33.105 \cite{3gpp-ts-33.105}. Instead, it utilizes a part of the cryptographic functions such as $\mathsf{SHA256}$, $\mathsf{f1}$, $\mathsf{f2}$, $\mathsf{f3}$, and $\mathsf{f4}$. We assume that these cryptographic functions are publicly available and provide confidentiality and integrity of their inputs. It is important to note that our protocols do not utilize the cryptographic functions $\mathsf{f5}$, $\mathsf{f1^*}$, and $\mathsf{f5^*}$ from the 5G-AKA protocol. Additionally, the Elliptic Curve Integrated Encryption Scheme (ECIES) \cite{Shoup2001} is employed as a secure public-key encryption. These assumptions regarding cryptographic functions are aligned with the requirements outlined in the 3GPP standards.

\paragraph*{Assumption on Components} Aligned with 3GPP standards, our threat model assumes the possibility of certain SNs and HNs being compromised. The long-term secret key ($\mathsf{k}$) of honest subscribers always remains secure, and the honest subscribers are capable of protecting their anchor keys ($\mathsf{K_{SEAF}}$). However, in proving our PFS property, we consider a scenario where the attacker has access to both the long-term secret key ($\mathsf{k}$) of the subscribers and the private key ($\mathsf{sk_{HN}}$) of the HN. Please note that the long-term secret key $\mathtt{k}$ of the subscriber may occasionally need to be updated for security reasons. In such cases, the mobile network operator can employ an over-the-air (OTA) SIM provisioning process to update the subscribers' long-term secret keys. However, such mechanisms fall outside the scope of our protocol design.

\subsection{Security Requirements}
\label{subsec:security_goals}
The security requirements outlined in the 3GPP security specifications can be broadly categorized into three groups: privacy, secrecy, and authentication \cite{Wang2021}. We define all the necessary security requirements specified in the 3GPP security specifications, along with the additional security requirements outlined in \cite{Basin2018} and \cite{Cremers2019}, which are not fully specified by 3GPP.


\paragraph*{\textbf{Privacy}} According to the 3GPP TS 33.501 \cite{3GPP} standards, a subscriber's privacy requirements can be classified into three types: user identity confidentiality, user location confidentiality, and user untraceability. These privacy requirements can be met by safeguarding the secrecy of the $\mathrm{SUPI}$ and ensuring the subscriber's untraceability, as demonstrated in \cite{Basin2018}\footnote{Please note that our protocols do not require sequence numbers, unlike 5G-AKA. Therefore, the privacy requirement of the subscriber in our protocols is independent of the sequence number.}. Our protocols should support these privacy requirements in the presence of both passive and active attackers, as indicated in \cite{Basin2018}. Furthermore, Wang \emph{et al.} \cite{Wang2021} demonstrated that all mentioned privacy requirements can be achieved if our protocol supports the indistinguishability property. This property asserts that no attacker can distinguish between two subscribers. We summarize this property below:
\par
\emph{\textbf{Subscriber Indistinguishability}}: If there are two subscribers, denoted as UE1 and UE2, and an AKA session involves UE1 (or UE2), no active attacker can distinguish whether it is engaged with UE1 or UE2.

\paragraph*{\textbf{Secrecy}} In the 3GPP TS 33.501 \cite{3GPP}, in addition to the secrecy of the $\mathrm{SUPI}$ and long-term secret key ($\mathsf{k}$), there is also the requirement for the secrecy of the anchor keys ($\mathsf{K_{SEAF}}$). As demonstrated in \cite{Basin2018}, the 5G-AKA protocol does not provide PFS. Therefore, our protocol II aims to support PFS as well. We outline the two secrecy requirements below:
\par 
\emph{\textbf{Key Secrecy}}: The anchor key ($\mathsf{K_{SEAF}}$) must remain secret.
\par 
\emph{\textbf{Perfect Forward Secrecy}}: If the long-term secret key ($\mathsf{k}$) and the long-term private key ($\mathsf{sk_{HN}}$) of the HN are compromised, the attacker must not be able to compute any previously generated anchor keys ($\mathsf{K_{SEAF}}$).

\paragraph*{\textbf{Authentication}} Our protocols also aim to provide all the authentication requirements specified in \cite{Basin2018} and \cite{Cremers2019}, which are formulated from the 3GPP TS 33.501 \cite{3GPP} standards. In \cite{Basin2018} and \cite{Cremers2019}, the authors used Lowe's taxonomy \cite{Lowe1997} to represent the authentication properties. We define the following definitions based on Lowe's taxonomy \cite{Lowe1997}:
\begin{itemize}
    \item \emph{Weak Agreement}: Weak agreement occurs when participant A (acting as the initiator) completes a protocol run with participant B, and participant B must have previously participated in the protocol with participant A.

    \item \emph{Non-Injective Agreement}: Non-injective agreement means that participant A completes a protocol run with participant B, and participant B must have previously participated in the protocol with participant A, with both A and B agreeing on the contents of all the messages exchanged.  

    \item \emph{Injective Agreement}: Injective agreement is a stronger version of non-injective agreement, requiring a one-to-one correspondence between the runs of A and B. It ensures that each run of A corresponds to a unique run of B, with both participants agreeing on the data or secrets involved.
\end{itemize}
We list the required authentication properties of our protocols below:
\par 
\emph{Agreement between UE and SN}: The subscriber and SN must both obtain injective agreement on $\mathsf{K_{SEAF}}$ and weak agreement with each other.
\par 
\emph{Agreement between UE and HN}: The subscriber and HN must achieve injective agreement regarding $\mathsf{K_{SEAF}}$ and establish a weak agreement with each other. Additionally, both the subscriber and the HN must achieve a non-injective agreement regarding $\mathsf{ID_{SN}}$ and $\mathsf{SUPI}$ with each other.
\par 

\emph{Agreement between SN and HN}: The SN and HN must both achieve injective agreement on $\mathsf{K_{SEAF}}$ and weak agreement with each other. The SN must also achieve a non-injective agreement on SUPI with HN.

\begin{figure*}[!t]
    \centering
	\fbox{\scalebox{6}{\includegraphics[width=2.5cm, height=1.9cm]{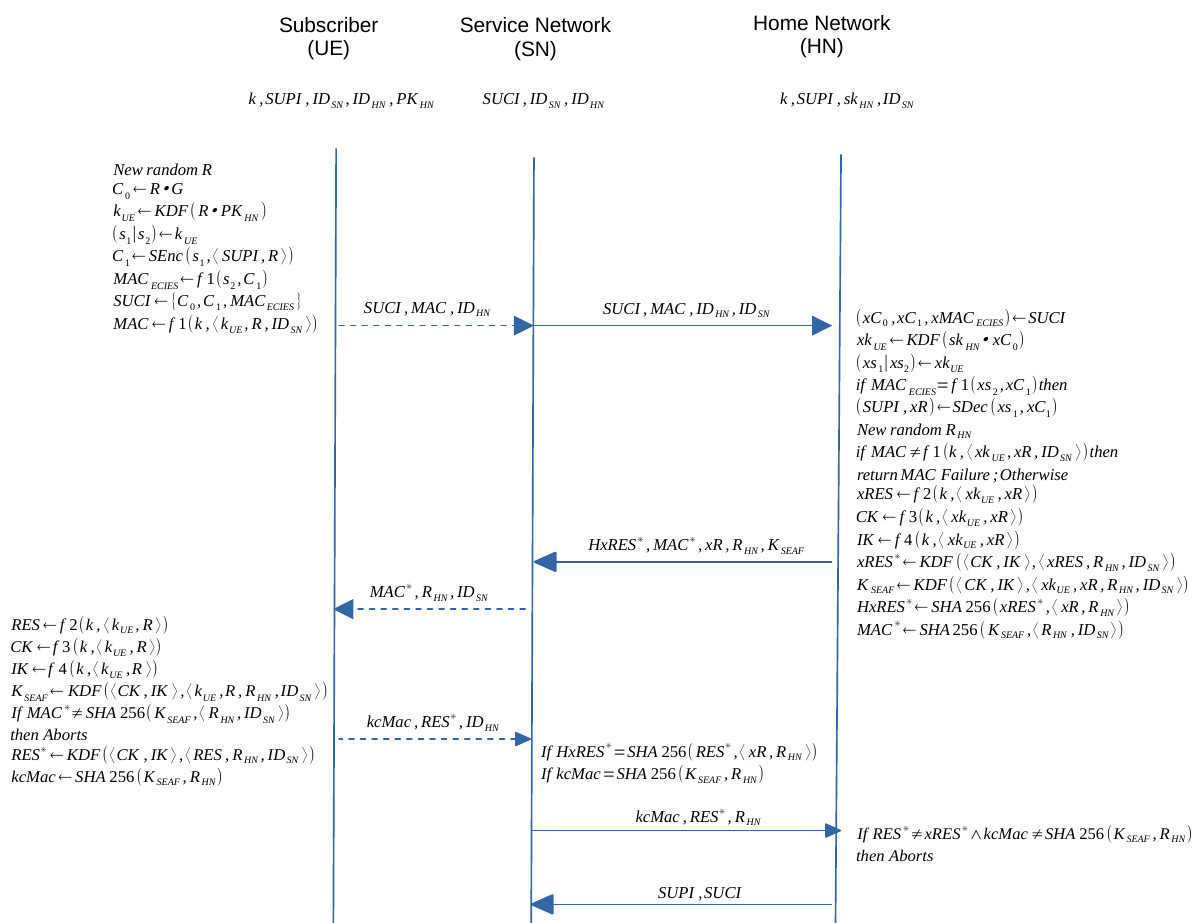}}}
	\caption{A high-level overview of our protocol I, where dotted and solid arrows represent open channels and authenticated secure channels, respectively.}		
    \label{fig:our-scheme}
 \end{figure*}	

\section{Our Proposed Protocol}
\label{sec:our_scheme}
Our protocol aims to streamline the AKA mechanism in 5G by addressing the limitations of the existing 5G-AKA protocol, such as enhancing subscriber privacy, ensuring PFS, and introducing an efficient replay attack prevention method. The objective is to meet all the security requirements outlined in the 3GPP standards and accommodate additional crucial security needs that are not fully specified in the standards, as highlighted in \cite{Basin2018}. This section begins with an overview of Protocol I. Then, we provide a detailed description of the protocol in Section \ref{sec:main_construction}. In Section \ref{sec:PFS}, we present Protocol II, an extension of Protocol I to support PFS.


\subsection{Overview}
As pointed out in \cite{Basin2018}, the 5G-AKA protocol exhibits vulnerabilities against active attackers. Several attack scenarios aimed at compromising subscriber privacy have been illustrated \cite{Basin2018}, \cite{Fouque2016}, \cite{Arapinis2012}, \cite{Borgaonkar2019} particularly focusing on capturing and replaying messages from targeted subscribers. The idea used by the attacks is to distinguish the responses elicited from the targeted subscriber and other subscribers to the same replayed messages. This enables attackers to differentiate them, thus breaching the privacy of the targeted subscriber. Appendix \ref{appendix:5G-AKA-Privacy-Issue} presents a brief overview of the common types of privacy-related attacks in the 5G-AKA protocol and the methods used to carry out these attacks.
\par 
The primary vulnerability in 5G-AKA lies in conducting two security checks at the subscriber side for the HN's challenge message, as highlighted in \cite{Wang2021}. Initially, the subscriber verifies the authenticity and integrity of the HN's random challenge by checking the $\mathsf{MAC}$. Subsequently, the subscriber assesses the freshness of the HN's challenge by comparing the subscriber's sequence number $\mathsf{SQN_{UE}}$ with the HN's sequence number $\mathsf{SQN_{HN}}$. For instance, in the \emph{Encrypted SUPI Replay Attack} \cite{Fouque2016}, \cite{Arapinis2012}, the attacker captures the $\mathrm{SUCI}$ of the targeted subscriber and replays it later. Consequently, the targeted subscriber responds to the HN's challenge, whereas other subscribers return \textsc{MAC\_Failure} responses. Similarly, in the \emph{Failure Message Linkability Attack} \cite{Arapinis2012}, \cite{Basin2018}, the attacker captures $R$ and $\mathrm{AUTN}$ sent by the HN to the targeted subscriber and replays them later. As a result, the targeted subscriber successfully verifies the $\mathrm{MAC}$ but fails to confirm freshness, responding with a \textsc{Sync\_Failure} message. Conversely, all other subscribers fail the $\mathrm{MAC}$ verification and respond with \textsc{MAC\_Failure} messages. The authors of \cite{Borgaonkar2019} demonstrated another type of privacy-related attack called the \emph{Sequence Number Inference Attack} in the 5G-AKA protocol by capturing $R$ and $\mathrm{AUTN}$ and replaying them later. 
\par 
We observe two primary reasons in the 5G-AKA protocol for its lack of resistance to active attackers. Firstly, there is a lack of binding between the subscriber's initial authentication request (i.e., $\mathrm{SUCI}$) and the corresponding HN's challenge ($R, \mathrm{AUTN}$). The subscriber cannot ascertain whether the HN's challenge is linked to its most recent authentication request. Secondly, the freshness check (or prevention of replay attacks) relies on a sequence number maintained by both the subscriber and HN. 

\par 
Our protocol I is designed to address the above-mentioned issues in the 5G-AKA protocol while ensuring that all essential security requirements outlined in Section \ref{sec:system_threat_model_security_requirements} are met. Unlike the 5G-AKA protocol, which relies solely on the HN sending the challenge and the subscriber's corresponding response, our protocol I incorporates an additional challenge message from the subscriber to the HN during the initial authentication request. These challenge and response messages are mutually bound using random numbers, eliminating the need for a sequence number-based replay attack prevention mechanism for freshness checking. This, in turn, also reduces the complexity of our protocol compared to the 5G-AKA protocol. Next, we provide a detailed description of Protocol~I. 

\subsection{Main Construction}
\label{sec:main_construction}
Figure \ref{fig:our-scheme} provides a high-level overview of our protocol I. It is divided into three phases: \emph{Initiation \& UE's Challenge}, \emph{HN's Challenge \& Response} and \emph{UE's Response \& Key Confirmation}, as will be described shortly. Our protocol I uses the cryptographic functions such as, $f1, f2, f3, f4, \mathrm{KDF}$ and $\mathrm{SHA256}$, as documented in 3GPP TS 33.501 \cite{3GPP}.
\par 
{\textbf{Initiation \& UE's Challenge}} The subscriber (UE) initiates this phase when the SN triggers authentication. In this phase, the subscriber primarily sends two important messages to the HN: the encrypted $\mathrm{SUPI}$, which is $\mathrm{SUCI}$, and a challenge $R$. The subscriber mainly performs two tasks, as outlined below.
\par 
Firstly, the subscriber chooses a random challenge $R$ and generates a randomized encrypted $\mathsf{SUPI}$ for privacy protection using ECIES, as outlined in Appendix \ref{sec:ECIES}. The ECIES key encapsulation mechanism $\mathsf{Encap_{ECIES}}$ generates a shared secret key $\mathsf{k_{UE}}$ and an ephemeral public-key component $C_0$. Concurrently, the data encryption mechanism of ECIES produces an encrypted component $C_1$ for the $\mathsf{SUPI}$ and the random challenge $R$ using symmetric key encryption. The $\mathsf{s_1}$ portion of the shared secret key $\mathsf{k_{UE}}$ is used as the encryption key, while the remaining $\mathsf{s_2}$ portion creates a tag (i.e., message authentication code) $\mathsf{MAC_{ECIES}}$ for authenticity and integrity verification of $C_1$.
\par 
Secondly, a message authentication code $\mathsf{MAC}$ is computed for the random challenge $R$. This will verify the authenticity and integrity of the challenge $R$.

\par 
Finally, the subscriber sends the tuple $\mathrm{SUCI}= \left<C_0, C_1, \mathrm{MAC_{ECIES}}\right>$, $\mathrm{MAC}, \mathrm{ID_{HN}}$ to the SN. Then, the SN forwards the tuple $\left<\mathrm{SUCI}, \mathrm{MAC}, \mathrm{ID_{HN}}, \mathrm{ID_{SN}}\right>$ to the HN.

\par 
\textbf{HN's Challenge \& Response} In this phase, the HN performs broadly two tasks once the HN receives the tuple $\left<\mathrm{SUCI}, \mathrm{MAC}, \mathrm{ID_{HN}}, \mathrm{ID_{SN}}\right>$ from a subscriber.
\par 
Firstly, the HN recovers the plaintext $\mathrm{SUPI}$ using its private key $\mathsf{sk_{HN}}$ from $\mathrm{SUCI}$ to retrieve the long-term secret key $\mathsf{k}$ associated with the subscriber from its database. Following the decapsulation operation of ECIES using its private key $\mathsf{sk_{HN}}$ and the ephemeral public-key component $C_0$ of $\mathrm{SUCI}$, the HN acquires the shared secret key $\mathsf{xk_{ue}}$. Subsequently, it decrypts $C_1$ using $\mathsf{xs_1}$ (a component of the shared secret $\mathsf{xk_{ue}}$) after successfully verifying the tag (i.e., $\mathsf{MAC_{ECIES}}$) using $\mathsf{xs_2}$ (the second component of the shared secret $\mathsf{xk_{ue}}$) and $C_1$, obtaining the plaintext $\mathrm{SUPI}$ and the subscriber's challenge~$xR$.

\par 

Secondly, the HN verifies the authenticity and integrity of the subscriber's challenge $xR$ by checking the $\mathrm{MAC}$. An unsuccessful verification results in a \textsc{MAC\_Failure}. Once the verification is successful, the HN chooses a random challenge for the subscriber $R_{HN}$ and computes $\mathrm{xRES}$, cipher key $\mathsf{CK}$, and integrity key $\mathsf{IK}$ using the long-term secret key $\mathsf{k}$ of the subscriber and the challenge $xR$. It then generates the expected response from the subscriber $\mathrm{xRES^*}$, and the anchor keys $\mathsf{K_{SEAF}}$. Next, the HN computes the hashed response $\mathrm{HxRES^*}$, which includes the expected response $\mathrm{xRES^*}$, the subscriber's challenge $xR$, and its challenge for the subscriber $R_{HN}$. Additionally, the HN generates a message authentication code $\mathrm{MAC^*}$. The HN sends the tuple $\left<\mathrm{HxRES^*}, \mathrm{MAC^*}, xR, R_{HN}, \mathsf{K_{SEAF}}\right>$ to the SN. Note that $\mathrm{MAC^*}$ serves three important functions: it acts as the response from the HN to the subscriber's challenge, it provides authenticity and integrity verification for the HN's challenge $R_{HN}$, and it offers key confirmation for the anchor key $\mathsf{K_{SEAF}}$ to the subscriber.

\par 
Upon receiving the tuple $\big<\mathrm{HxRES^*}, \mathrm{MAC^*}, xR, R_{HN}, \mathsf{K_{SEAF}}\big>$, the SN retains $\big<\mathrm{HxRES^*}, xR, R_{HN}, \mathsf{K_{SEAF}}\big>$ and forwards the tuple $\left<\mathrm{MAC^*}, R_{HN}, \mathrm{ID_{SN}}\right>$ to the subscriber.

\par
\textbf{UE's Response \& Key Confirmation}: Upon receiving the tuple $\left<\mathrm{MAC^*}, R_{HN}, \mathrm{ID_{SN}}\right>$, the subscriber computes the response $\mathsf{RES}$, cipher key $\mathsf{CK}$, and integrity key $\mathsf{IK}$ using $f2$, $f3$, and $f4$ respectively, along with the long-term secret key $\mathsf{k}$, shared secret key $\mathsf{k_{UE}}$, and the random challenge $R$ as inputs. The subscriber then computes the anchor keys $\mathsf{K_{SEAF}}$ and verifies the $\mathrm{MAC^*}$ using $\mathsf{K_{SEAF}}$, HN's challenge $R_{HN}$, and $\mathrm{ID_{SN}}$.
\par 
If the verification fails, the connection is aborted. If the verification is successful, it provides three functions: authenticity and integrity of the HN's challenge $R_{HN}$, confirmation of the anchor key $\mathsf{K_{SEAF}}$, and verification of the HN's response to its challenge $R$. Afterward, the subscriber generates its response $\mathrm{RES^*}$ and a message authentication code $\mathrm{kcMAC}$ using $\mathrm{SHA256}$, and sends the tuple $\left<\mathrm{kcMAC}, \mathrm{RES^*}, \mathrm{ID_{SN}}\right>$ to the SN. Note that $\mathrm{kcMAC}$ serves two purposes: verifying the subscriber's response to the HN's challenge $R_{HN}$ and confirming the anchor keys $\mathsf{K_{SEAF}}$ to the HN.

\par 
Upon receiving the response from the subscriber, SN first verifies $\mathrm{HxRES}$ with $\mathrm{RES^*}, xR$ and $R_{HN}$. It also verifies $\mathrm{kcMAC}$ with the anchor keys $\mathsf{K_{SEAF}}$ and $R_{HN}$. A successful verification indicates authentication of the subscriber at the SN and anchor key $\mathsf{K_{SEAF}}$ confirmation. Next, the SN forwards the tuple $\big<\mathsf{kcMAC}, \mathrm{RES^*}, R_{HN}\big>$ to the HN, where $R_{HN}$ is sent for the binding purposes. 
\par 
Upon receiving the tuple $\left<\mathsf{kcMAC}, \mathrm{RES^*}, R_{HN}\right>$, the HN verifies $\mathrm{RES^*}$ with $\mathrm{xRES^*}$ and also $\mathsf{kcMAC}$ using $\mathsf{K_{SEAF}}$ and $R_{HN}$. A successful verification indicates mutual authentication between the subscriber and the HN and confirms the anchor keys $\mathsf{K_{SEAF}}$. The HN then sends the tuple $\left<\mathrm{SUPI}, \mathrm{SUCI}\right>$ to the SN, where $\mathrm{SUCI}$ is used for binding purposes. Once the SN receives $\mathrm{SUPI}$, the authentication process is concluded, and the subscriber can use the anchor key $\mathsf{K_{SEAF}}$ to access network services.

\begin{figure*}[!t]
    \centering
	\fbox{\scalebox{7}{\includegraphics[width=2.1cm, height=1.8cm]{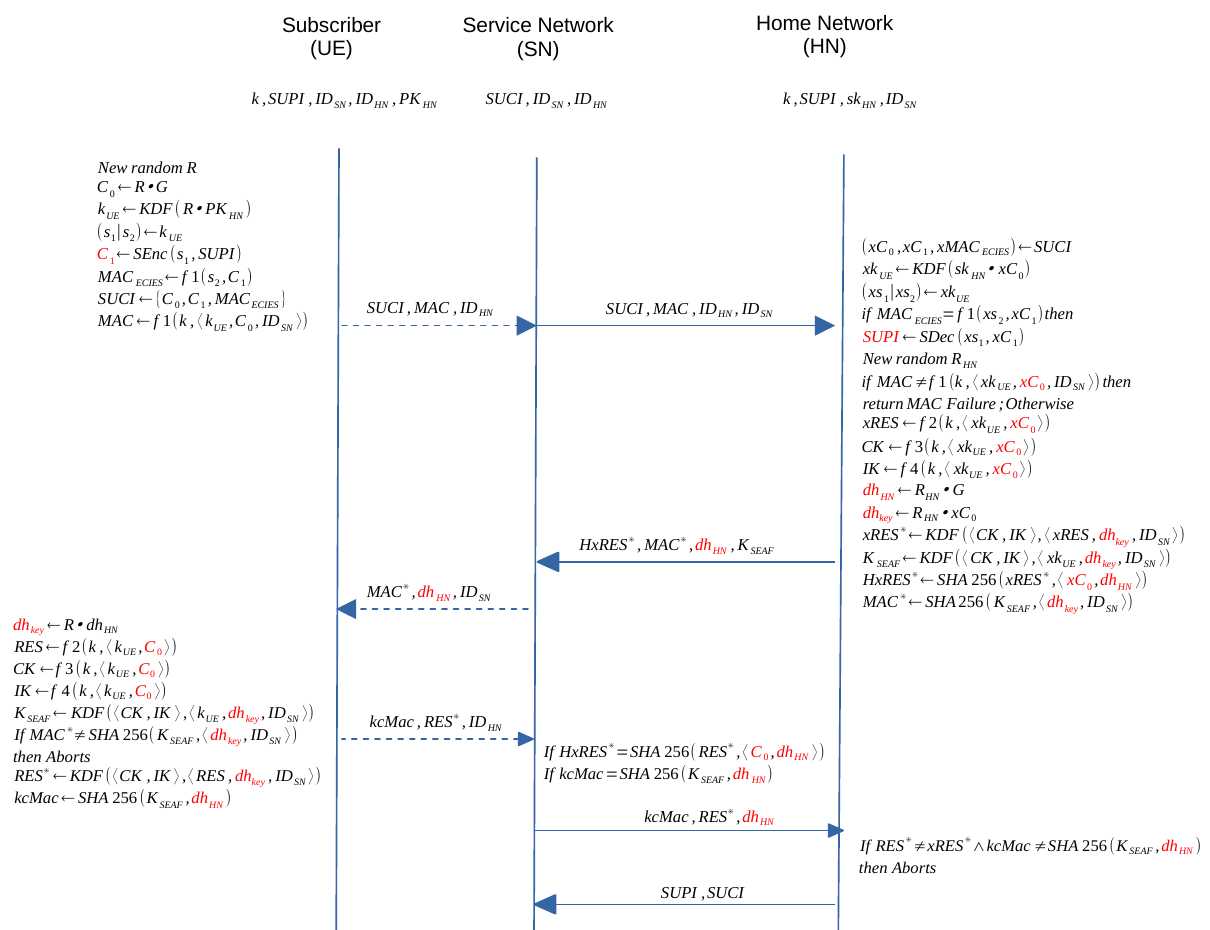}}}
	\caption{A high-level overview of our protocol II, where dotted and solid arrows represent open channels and authenticated secure channels, respectively.}		
    \label{fig:our-scheme-fs}
 \end{figure*}

\subsection{Extension for Perfect Forward Secrecy (PFS)}
\label{sec:PFS}
In this section, we present Protocol II, an extension of Protocol I, designed to support PFS with minimal additional computational overhead. Please note that Protocol I is a foundational enhancement to the 5G-AKA protocol, addressing key limitations without relying on PFS. It replaces sequence number-based replay protection with a stateless challenge-response mechanism, simplifying synchronization and reducing overhead. This design also improves resistance to active attacks by binding authentication messages between the subscriber and the home network. Crucially, Protocol I remains compatible with existing systems, making it practical for deployment in current 5G networks.
\par 
Protocol II builds on this foundation by introducing PFS, which protects past session keys even if long-term secrets are later compromised, which is a growing concern in the face of persistent and sophisticated adversaries. This layered design ensures that Protocol I is practical and deployable on its own, while Protocol II offers additional security for more demanding threat models. 
\par 
To achieve PFS, our previously described Protocol I requires some modifications. Figure \ref{fig:our-scheme-fs} shows our modified protocol, with the highlighted portions indicating the new changes. The core addition in our modified protocol II is the introduction of an ephemeral DH key exchange between the subscriber and the HN. Please note that our modified protocol II supports all the security requirements of our protocol I, which are confirmed by our formal analysis.
\par 
 As we aim to maintain USIM compatibility without introducing new cryptographic operations, our core approach is to utilize the ECIES ephemeral public key $C_0$ as DH key material for computing a shared DH key between the subscriber and the HN. In Figure \ref{fig:our-scheme-fs}, it is shown that the HN employs the ephemeral public key $xC_0$ (i.e., $C_0$) to calculate a DH key $\mathsf{dh_{key}}$, using its random challenge $R_{HN}$ as the secret key. Furthermore, the HN computes the DH key material $dh_{HN}$ for the subscriber using $R_{HN}$, enabling the subscriber to compute the same DH key $\mathsf{dh_{key}}$ on its end. Subsequently, both the subscriber and the HN derive the anchor keys (i.e., $\mathsf{K_{SEAF}}$) using the $\mathsf{dh_{key}}$, $\mathsf{k_{UE}}$, and $\mathrm{ID_{SN}}$. To avoid redundancy, we refrain from reiterating the entire protocol, as the remaining concept aligns with our protocol I.

\begin{table}[h]
\caption{Security Requirements Achieved by Our Protocols}
\label{table:verification_results}
\centering
\begin{tabular}{|l|ll|ll|ll|}
\hline
Point of View  & \multicolumn{2}{c|}{UE}      & \multicolumn{2}{c|}{SN}      & \multicolumn{2}{c|}{HN}      \\ \hline
Partner        & \multicolumn{1}{l|}{SN} & HN & \multicolumn{1}{l|}{UE} & HN & \multicolumn{1}{l|}{UE} & SN \\ \hline
Weak agreement & \multicolumn{1}{l|}{\cmark}   &  \cmark  & \multicolumn{1}{l|}{\cmark}   &  \cmark  & \multicolumn{1}{l|}{\cmark}   &  \cmark  \\ 
Agreement on $\mathsf{K_{SEAF}}$ & \multicolumn{1}{l|}{I}   &  I  & \multicolumn{1}{l|}{I}   & I   & \multicolumn{1}{l|}{I}   & I   \\ 
Agreement on $\mathsf{ID_{SN}}$ & \multicolumn{1}{l|}{NI}   &  NI  & \multicolumn{1}{l|}{NI}   &  NI  & \multicolumn{1}{l|}{NI}   &  NI  \\ 
Agreement on $\mathsf{SUPI}$ & \multicolumn{1}{l|}{NI}   &  NI & \multicolumn{1}{l|}{NI}   &  NI  & \multicolumn{1}{l|}{NI}   &  NI  \\ \hline
Secrecy on $\mathsf{K_{SEAF}}$ & \multicolumn{2}{c|}{\cmark}  & \multicolumn{2}{c|}{\cmark}  & \multicolumn{2}{c|}{\cmark}  \\ \hline
Secrecy on $\mathsf{SUPI}$ & \multicolumn{2}{c|}{\cmark}  & \multicolumn{2}{c|}{\cmark}  & \multicolumn{2}{c|}{\cmark}  \\ \hline
PFS$^*$ & \multicolumn{6}{c|}{\cmark}     \\\hline
Indistinguishability & \multicolumn{6}{c|}{\cmark}     \\\hline
\end{tabular}
\\
\scriptsize {I: Injective agreement; wa: Weak agreement; NI: Non-injective agreement; \cmark: Property supported; $^*$: only our protocol II support PFS.} 
\end{table}


\begin{table*}[!t]
\centering

\begin{tabular}{cccccc}\hline \hline
\multicolumn{2}{c}{} & 5G-AKA \cite{3GPP} & 5G-AKA$'$ \cite{Wang2021} & Our Protocol I & Our Protocol II \\\hline \hline
\multicolumn{2}{c}{Resynchronization} & Yes & Yes & No & No \\\hline
\multirow{2}{*}{Messages} & Resync &  $13$& $13$ & \multirow{2}{*}{} NA & \multirow{2}{*}{} NA \\
 & No Resync & $9^*$ & $9^*$ & $7$ & $7$ \\ \hline
\end{tabular}
\\
\scriptsize{$^*$5G-AKA and 5G-AKA$'$ require $7$ message interactions between the parties for mutual authentication and an additional $2$ messages for the anchor key $\mathsf{K_{SEAF}}$ confirmation between the subscriber and the SN; NA: Not Applicable. }
\caption{ Communication overhead comparison between our protocols (Protocols I and II) and 5G-AKA \cite{3GPP} and 5G-AKA$'$} 
\label{table:message-comparison}
\end{table*}


\begin{table*}[!t]
\centering
\tiny
\begin{tabular}{|p{1cm}|p{1.5cm}|p{.5cm}|p{1.5cm}|p{1.5cm}|p{.5cm}|p{1.5cm}|p{1.5cm}|p{.5cm}|p{1.5cm}|} \hline 
\multirow{2}{*}{} & \multicolumn{3}{c|}{Case 1} & \multicolumn{3}{c|}{Case 2} & \multicolumn{3}{c|}{Case 3} \\\hline
 & UE & SN & HN & UE & SN & HN & UE & SN & HN \\ \hline 
5G-AKA \cite{3GPP}& $8T_h+2 T_m+ T_{enc}+T_{xor}+ T_{add}+ T_{prf}$ & $T_h$ & $9T_h+T_m+T_{dec}+T_{xor}+ T_{add}+ T_{prf}$ & $6T_h+ 2T_m+ T_{enc}+ 2T_{xor}+ T_{prf}$ & $T_h$ & $11T_h+ T_m+ T_{dec}+ 2T_{xor}+ 2T_{add}+ T_{prf}$ & $4T_h+ 2T_m+ T_{enc}+T_{xor}+ T_{prf}$ & $T_h$ & $9T_h+ T_m+ T_{dec}+ T_{xor}+ T_{add}+ T_{prf}$ \vspace{.15cm}\\\hline
5G-AKA$'$ \cite{Wang2021}& $8T_h+ 2T_m+ T_{enc}+T_{xor}+ T_{add}+ T_{prf}+ T_{dec}$ & $T_{h}$ & $9T_h+ T_m+ T_{enc}+ T_{xor}+ T_{add}+ T_{prf}+ T_{dec}$ & $6T_h+ 2T_m+ T_{enc}+ 2T_{xor}+ T_{prf}+ T_{dec}$ & $T_{h}$  & $11T_h+ T_m+ T_{enc}+ 2T_{xor}+ 2T_{add}+ T_{prf}+ T_{dec}$ & $4T_h+ 2T_m+ T_{enc}+ T_{xor}+ T_{prf}+ T_{dec}$ & $T_{h}$  &  $9T_h+ T_m+ T_{enc}+ T_{xor}+ T_{add}+ T_{prf}+ T_{dec}$ \vspace{.15cm}\\\hline
Our Protocol I & $10T_h+ 2T_m+ T_{enc}+ T_{prf}$ & $2T_{h}$ & $11T_h+ T_m+ T_{dec}+ T_{prf}$ & NA&NA& NA&$8T_h+ 2T_m+ T_{enc}+ T_{prf}$ & $2T_{h}$& $10T_h+ T_m+ T_{dec}+ T_{prf}$ \vspace{.15cm}\\\hline
Our Protocol II & $10T_h+ 3T_m+ T_{enc}+ T_{prf}$ & $2T_{h}$ & $11T_h+ 3T_m+ T_{dec}+ T_{prf}$ & NA&NA& NA & $8T_h+ 3T_m+ T_{enc}+ T_{prf}$& $2T_{h}$& $10T_h+ 3T_m+ T_{dec}+ T_{prf}$\\\hline
\end{tabular}
\caption{Theoretical Computation Cost Comparison between Our Protocol I, Protocol II, 5G-AKA \cite{3GPP} and 5G-AKA$'$ \cite{Wang2021}} 
\label{table:performamce-evaluation}
\end{table*}

\begin{table*}[!t]
\centering
\tiny
\begin{tabular}{|c|p{1.5cm}p{.5cm}p{1.5cm}|p{1.5cm}p{.5cm}p{1.5cm}|p{1.5cm}p{.5cm}p{1.5cm}|} \hline
\multirow{2}{*}{} & \multicolumn{3}{c|}{Case 1} & \multicolumn{3}{c|}{Case 2} & \multicolumn{3}{c|}{Case 3} \\\hline
 & UE & SN & HN & UE & SN & HN & UE & SN & HN \\ \hline 
5G-AKA \cite{3GPP}& $39609.97$ & $22.57$ & $38529.26$ & $31340.20$ & $21.14$ & $38614.81$ & $18600.93$ & $21.42$& $39314.69$\\
5G-AKA$'$ \cite{Wang2021} & $40539.53$ & $23.49$ & $39572.49$ & $32023.00$ & $24.29$ & $39820.31$ & $19107.45$ & $23.05$ & $39450.83$\\
Our Protocol I & $39631.14$& $32.71$ & $40627.80$ &NA &NA&NA & $28835.33$& $32.28$&$40436.40$\\
Our Protocol II & $39664.36$& $30.96$ & $46579.51$&NA &NA&NA & $28857.77$ &$31.73$& $46062.19$\\\hline
\end{tabular}
\caption{Computation Time (in microseconds) Comparison between Our Protocol I, Protocol II, 5G-AKA \cite{3GPP} and 5G-AKA$'$ \cite{Wang2021}} 
\label{table:computation-time}
\vspace{-.5cm}
\end{table*}

\section{Formal Security Analysis}
\label{sec:formal_security_analysis}
 ProVerif is a state-of-the-art symbolic model-based automated formal analysis tool that uses applied $\pi$-calculus syntax \cite{proverif}. It operates under the \emph{Dolev-Yao Attack Model} \cite{Dolev1983}, which grants the attacker the ability to read, modify, delete, and forge packets, as well as inject them into the public communication channel. ProVerif evaluates whether the designed protocol meets the defined security objectives within this attack model. When an attack is detected, ProVerif provides a comprehensive description of the steps involved in the attack. ProVerif is ideal for stateless protocols like ours and for verifying a wide range of security properties under the Dolev-Yao model, including indistinguishability properties\footnote{5G-AKA is a stateful protocol due to its sequence number-based replay attack prevention mechanism, which makes it unsuitable for modeling using ProVerif. Consequently, most existing formal analyses of the 5G-AKA protocol are based on Tamarin \cite{Tamarin}, which supports stateful protocols.}. For our analysis, we use ProVerif version $2.05$ \cite{proverif}. Our modeling codes are available in \url{https://anonymous.4open.science/r/AKA-5G-ProVerif-9378/}.
\subsection{Modeling Choices}
In formal verification, modeling choices play a crucial role in defining the behavior, assumptions, and properties of the system to facilitate formal analysis. These choices abstractly represent the system or protocol under analysis. In this section, we will briefly outline the modeling choices made for our protocols to provide a better understanding of our modeled system.
\paragraph*{Architecture}
We consider three roles: subscribers, SNs, and HNs. Each role can have an unbounded number of instances. The communication channels between the subscriber and SN, and between the SN and HN, are considered open (or insecure) and authenticated private (or secure) channels, respectively, as mentioned in Section \ref{sec:system_threat_model_security_requirements}. 
\paragraph*{Modeling Cryptographic Primitives} 
We model symmetric key-based encryption and decryption operations using the constructor $\mathrm{senc}$ for encryption and the destructor $\mathrm{sdec}$ for decryption. The $\mathrm{senc}$ constructor takes two arguments: a message $m$ of type bitstring and a key $n$ of type bitstring, and returns the encrypted message. The $\mathrm{sdec}$ destructor takes an encrypted message produced by $\mathrm{senc}$ and the corresponding key $n$, and returns the original message $m$. Additionally, cryptographic hash functions such as $f1$, $f2$, $f3$, $f4$, $\mathrm{KDF}$, and $\mathrm{SHA256}$ are also modeled as constructors. We model the modular exponentiation operation (i.e., $g^x$) using the constructor $\mathrm{exp}$. This constructor takes two arguments of type $G$ and $\mathrm{exponent}$. For modeling the DH Key Exchange, we utilize the \emph{equation} concept in ProVerif. The DH Key Exchange is represented by the equation:   
\begin{align*}
    &\mathrm{const\hspace{.1cm} g:G\hspace{.1cm} [data].}\\
	&\mathrm{fun\hspace{.1cm} exp(G, exponent):G.}\\
	&\mathrm{equation\hspace{.1cm} forall\hspace{.1cm} x:exponent, y: exponent;}\\
    &\mathrm{exp(exp(g,x),y)= exp(exp(g,y),x).}
\end{align*}

\paragraph*{Security Goals Modeling}
Here, we provide a brief description of how the security requirements mentioned in Section \ref{subsec:security_goals} are modeled.
\paragraph*{Authentication} ProVerif provides correspondence assertions to capture the authentication or relationship between parties or events. We use both basic and injective correspondence assertions to capture non-injective (which also covers weak agreement) and injective agreements between parties, respectively. If the specified events occur in the correct sequence and the parameters remain consistent, the corresponding attribute is validated.
\par 
The following query demonstrates that if party A executes event $e1$ with parameter $x$, then party also performs event $e2$ using the same parameter. This implies that, from party A's perspective, party B has achieved non-injective correspondence with A concerning parameter $x$. An example of such non-injective correspondence is shown below: 
 \begin{align*}
 &\mathrm{query\hspace{.1cm}x:bitstring, t1, t2: time;} \\&\mathrm{event\hspace{.1cm} (e1(x)@t1)}
		\mathrm{==>\hspace{.1cm} (event(e2(x))@t2\hspace{.1cm} \&\&\hspace{.1cm} t1>t2}).
\end{align*}
The inclusion of the temporal parameters $t1, t2$ further refines the query by associating each event with its occurrence time, thereby enabling precise tracking of causality and potential attack vectors. Similarly, an example of injective correspondence is shown below:
\begin{align*}
&\mathrm{query\hspace{.1cm}x:bitstring, t1, t2, t3, t4, t5: time;} \\&\mathrm{inj-event\hspace{.1cm} (e1(x)@t1)}
		\mathrm{==>}\\&\mathrm{(inj-event(e2(x))@t2\hspace{.1cm} \&\&\hspace{.1cm} t1>t2)}.
\end{align*}
This injective correspondence implies that there is a one-to-one relationship between the number of protocol runs performed by each participant. The injective correspondence assertion asserts that for each occurrence of $\mathrm{event} \hspace{.1cm} e1(x)$, there is a distinct earlier occurrence of $\mathrm{event} \hspace{.1cm}e2(x)$.

\paragraph*{Secrecy} We leverage the reachability property of ProVerif to prove our secrecy claims. The reachability property enables the ProVerif tool to automatically search for any terms accessible to an attacker. Therefore, if a term, e.g., $x$, is accessible to an attacker, ProVerif can help identify the attack vector. We use the following queries to check whether a term $x$ is accessible to the attacker, where the term $x$ could be the anchor key $\mathsf{K_{SEAF}}$, long-term secret key $\mathsf{k}$, subscriber identity $\mathsf{SUPI}$, and HN's private key $\mathsf{sk_{HN}}$. For instance, the following query allows ProVerif to verify whether $x$ remains confidential:
\begin{align*}
   & \mathrm{query\hspace{.1cm} x: bitstring, t1, t2, t3, t4, t5: time;} \\
	&\mathrm{((event(e1(x))@t1\hspace{.1cm} \&\&\hspace{.1cm} attacker(x)@t2)}\hspace{.1cm} \mathrm{==> false}).
\end{align*}

\paragraph*{Modeling Active Attacker and Subscribers Privacy} In ProVerif, achieving the indistinguishability property entails ensuring that an active attacker cannot distinguish two different processes in the protocol execution. This is typically accomplished by designing the protocol in a manner that prevents any information leaked to the active attacker from providing an advantage in distinguishing between these processes. The concept of indistinguishability is represented using observational equivalence in ProVerif.
\par
To model observational equivalence between two processes in our model, we utilize the construct $\mathsf{Choice[M, M']}$, which provides a single \emph{biprocess} encoding both processes. The $\mathsf{Choice[M, M']}$ encapsulates the terms that differ between the two processes: one process uses the first component, $\mathsf{M}$, while the other process uses the second one, $\mathsf{M'}$. Our analysis reveals that these two processes exhibit equivalence, implying they possess identical structures and differ only in the selection of terms. This finding suggests that an active attacker cannot distinguish between these distinct processes during protocol execution. Hence, both of our protocols support protection against active attackers.

\subsection{Formal Verification Results}
Table \ref{table:verification_results} summarizes the security requirements supported by our protocols using ProVerif, as outlined in Section \ref{sec:system_threat_model_security_requirements}. Specifically, both protocols achieve injective agreement on $\mathsf{K_{SEAF}}$ for each pair of parties. Furthermore, they achieve non-injective agreement on $\mathsf{ID_{SN}}$ between the subscriber UE and HN, while the SN achieves non-injective agreement on $\mathsf{SUPI}$ with the HN. Both protocols also ensure the secrecy of the anchor key $\mathsf{K_{SEAF}}$ and subscriber's identity $\mathsf{SUPI}$. Additionally, Protocol II supports PFS. Our security analysis further demonstrates that both protocols effectively protect against active attackers (indistinguishability). 

\section{Performance Analysis}
\label{sec:performance-analysis}
In this section, we present a detailed comparison of our protocols with the 3GPP standardized 5G-AKA protocol \cite{3GPP} and its improved version 5G-AKA$'$ (USENIX'21) \cite{Wang2021}. We start with a theoretical comparison, followed by experimental results. 

\subsection{Theoretical Comparison}
\label{sec:theoritical-comparison}
This section compares theoretically our protocols with the 3GPP standardized 5G-AKA and 5G-AKA$'$. 
\par 
Table \ref{table:message-comparison} shows a comparison between our protocols with others in terms of the number of messages exchanged between entities in each protocol. Since neither of our protocols uses sequence numbers, they do not require a sequence number resynchronization phase. As a result, our protocols can be completed with only seven messages exchanged among the subscriber, SN, and HN. In contrast, 5G-AKA and 5G-AKA$'$ require $13$ and $9$ messages, respectively, when sequence number synchronization is necessary and when it is not. It is evident that our protocols require fewer message exchanges between entities, thereby reducing overall communication overhead. It is important to note that, compared to the 5G-AKA protocol, our protocols introduce an additional $\mathsf{MAC}$ parameter, which is transmitted from the subscriber to the HN via the SN along with the $\mathsf{SUCI}$. As detailed in Section \ref{sec:main_construction}, this addition enhances our protocols by avoiding the inefficient sequence number-based replay attack prevention mechanism used in 5G-AKA and by making them more resistant to active attackers. Additionally, our protocols introduce the $\mathsf{kcMAC}$ parameter, which is sent from the subscriber to the HN via the SN along with $\mathsf{RES^*}$. This parameter enables explicit key confirmation with both the SN and HN, a feature not present in the 5G-AKA protocol. Moreover, while the 5G-AKA protocol requires an additional parameter, $\mathsf{CONC}$, sent by the HN to the subscriber via the SN, our protocols do not require such a parameter.

\par 
Table \ref{table:performamce-evaluation} compares the computation costs between our protocols and 5G-AKA as well as 5G-AKA$'$. We evaluate the computation costs incurred by the subscriber, SN, and HN during the execution of each protocol. For 5G-AKA and 5G-AKA$'$, we consider three scenarios: successful authentication (Case 1), \textsc{Sync\_Failure} (Case 2), and \textsc{MAC\_Failure} (Case 3). For our protocols, we focus on Case 1 and Case 3, as Case 2 is not applicable\footnote{Our protocols do not require the sequence number resynchronization phase.}. In Cases 2 and 3, the computation costs reflect the effort required by the protocols to reach the \textsc{Sync\_Failure} and \textsc{MAC\_Failure} phases, respectively.
\par 
We denote $T_h$, $T_m$, $T_{enc}$, $T_{xor}$, $T_{add}$, $T_{prf}$, and $T_{dec}$ as the time required to perform one hash/KDF/MAC operation, elliptic curve scalar multiplication, encryption, XOR, addition, random number generation, and decryption operation, respectively. As shown in Table \ref{table:performamce-evaluation}, our protocols incur slightly higher computation costs for the subscriber, SN, and HN compared to the 5G-AKA protocol in Case 1. In Case 3, our protocols involve relatively more computation at both the subscriber and HN sides. However, as mentioned earlier, Case 3 is likely to occur less frequently. 

\subsection{Experimental Results}
\label{sec:experiment}
We implemented both of our protocols as well as 5G-AKA and 5G-AKA$'$ to compare their computation times at the subscriber, SN, and HN sides. The simulations were conducted on a GPU Laptop $11$ Enterprise ($64$-bit) machine with a $2.8$ GHz Intel (R) Xeon(R) E-2276M and $32$ GB of memory, using Microsoft Visual Studio $2022$. We utilized the \textsc{httplib} library \cite{httplib} to manage HTTP/HTTPS communications between the entities (i.e., subscriber, SN, and HN). All protocols were evaluated under the same security level. All our codes are available in \url{https://anonymous.4open.science/r/AKA-5G-E8B4/}.
\par
We used the Crypto++ cryptographic library \cite{crypto++} to implement ECIES with the \textsc{secp256r1} curve. The Encryptor.Encrypt() and Decryptor.Decrypt() interfaces were modified to support the export and import of shared keys derived by ECIES. Additionally, we employed $\mathrm{SHA256}$ with different prefixes as $\mathrm{f1, f2, f3, f4, f5, f1^*, f5^*}$. For key derivation and message authentication, we used Password-Based Key Derivation Function 2 (PBKDF2) and Hash-based Message Authentication Code (HMAC) for $\mathrm{KDF}$ and $\mathrm{HMAC}$, respectively. The computation time is measured using the \textsc{chrono} library provided by C++. 
\par 
Table \ref{table:computation-time} presents the actual computation times for the subscriber, SN, and HN to complete their respective operations, measured in microseconds. The results indicate that, in Case 1, our protocol I demonstrates a highly competitive performance, with a computation cost only $0.05\%$ higher than the 5G-AKA protocol and $2.29\%$ lower than the 5G-AKA$'$ protocol at the subscriber side. Our protocol II also performs well, with just $0.13\%$ additional computation time compared to the 5G-AKA protocol and $2.21\%$ less compared to the 5G-AKA$'$ protocol on the subscriber side.
\par 
Both of our protocols require an additional hash operation at the SN compared to the 5G-AKA and 5G-AKA$'$ protocols, resulting in a $48\%$ increase in computation time. While the 5G-AKA and 5G-AKA$'$ protocols involve a single hash operation at the SN, our protocols incorporate two hash operations. This enhancement significantly boosts the security of our protocols, justifying the increased computation time. At the HN side, our protocol I requires $5.45\%$ and $2.67\%$ more computation time than the 5G-AKA and 5G-AKA$'$ protocols, respectively. Our protocol II demonstrates a slightly higher increase, with $20.91\%$ and $17.71\%$ more computation time compared to the 5G-AKA and 5G-AKA$'$ protocols, respectively, reflecting its enhanced security benefits.

\par 
Our protocols avoid Case 2, which also provides better overall efficiency and reduces the computational overhead associated with sequence number resynchronization. This contributes to a more streamlined and effective authentication process, enhancing performance and reliability.
\par 
Case 3 occurs less frequently, typically due to an active attacker or a misconfiguration in the network or HN. Nevertheless, we evaluate the performance of our protocols under this scenario. At the subscriber side, our protocols incur approximately 55.02\% more computation time compared to the 5G-AKA protocol, and about 50\% more than the 5G-AKA$'$ protocol. On the HN side, Protocol I and Protocol II introduce increases of approximately 2.8\% and 17\%, respectively, over the 5G-AKA protocol. Compared to the 5G-AKA$'$ protocol, Protocol I requires 2.5\% more computation time, while Protocol II requires 16.7\% more. Given the rarity of this scenario, the additional computational overhead introduced by our protocols is considered acceptable, especially in light of the enhanced security guarantees they provide.

\section{Conclusion}
\label{sec:conclusion}
In this paper, we proposed AKA protocols for 5G that achieve all security goals specified in the 3GPP technical specification TS 33.501, along with important underspecified security objectives. We designed two protocols: the first provides all security guarantees except perfect forward secrecy, and its extended version also supports perfect forward secrecy with minimal computational overhead. Both protocols are compatible with existing SIM cards, resist both passive and active attacks, and may require only software modifications on the subscriber, SN, and HN. We verified the claimed security goals using ProVerif. Our implementation results and comparisons with the existing 5G-AKA and 5G-AKA$'$ protocols demonstrate that our protocols offer enhanced security while providing better or comparable computation and communication overhead, making them well-suited for 5G and beyond.
\par
While our simulation-based evaluation provides a comparative analysis of computational and communication overhead, it does not fully capture the operational dynamics of real-world 5G environments. Factors such as device heterogeneity, network variability, and mobility were not modeled. Additionally, integration with actual mobile network stacks and adversarial testing in live settings remains an open area for future work. We plan to extend our evaluation using real-world testbeds such as OpenAirInterface (OAI) to validate the practical effectiveness and robustness of our protocols under realistic conditions. Furthermore, while our protocol is designed to be compatible with existing USIM cards based on current 3GPP specifications, we acknowledge that this compatibility has not yet been experimentally verified. Additional testing is needed to confirm practical interoperability, which remains part of our future work.

\section*{Acknowledgment}
    This research paper is conducted under the 6G Security Research and Development Project, as led by the Commonwealth Scientific and Industrial Research Organisation (CSIRO) through funding appropriated by the Australian Government’s Department of Home Affairs. This paper does not reflect any Australian Government policy position. For more information regarding this Project, please refer to \url{https://research.csiro.au/6gsecurity/}.

\appendices
\section{Elliptic Curve Integrated Encryption Scheme (ECIES) \cite{Shoup2001}}
\label{sec:ECIES}
5G uses ECIES cryptographic primitive to protect the unique identity of the subscribers (please refer to 3GPP TS 33.501 \cite{3GPP}). ECIES is a hybrid encryption scheme based on public key cryptography, comprising a Key Encapsulation Mechanism (KEM) and a Data Encapsulation Mechanism (DEM). In ECIES, the KEM is used to establish shared keys between the sender and recipient through public key cryptography. Subsequently, a DEM encrypts and decrypts the actual payload using symmetric cryptography with the shared key. The ECIES consists of the following algorithms.

\paragraph*{\textsc{KeyGen} $((\mathsf{sk, pk})\leftarrow \mathsf{pp})$} It takes an elliptic curve domain parameters $\mathsf{pp}$ as input and outputs a private key $\mathsf{sk}$ and a public key $\mathsf{pk}$, where $\mathsf{sk}\in \mathbb{Z}_q^*$ (a multiplicative group of integer modulo $q$), $\mathsf{pk}= \mathsf{sk}\cdot g$, and $g$ is a generator of the chosen elliptic curve. 3GPP recommends two elliptic curves \textsc{Curve25519} and \textsc{secp256r1} \cite{3GPP}. 

\paragraph*{\textsc{Encap}$((C_0, \mathsf{k_s})\leftarrow \mathsf{pk})$} It takes the public key $\mathsf{pk}$ as input to generate an ephemeral private-public key pair $(r, R)$, where $r\in \mathbb{Z}_q^*$ and $R= r\cdot g$. It sets $C_0= R$ and computes a shared secret key $\mathsf{k_s}$, where $\mathsf{k_s}= \mathrm{KDF}(r\cdot \mathsf{pk})$.

\paragraph*{\textsc{Decap} $(\mathsf{k_s}\leftarrow (C_0, \mathsf{sk}))$} It takes $C_0$ and the private key $\mathsf{sk}$ as input and outputs the shared secret key $\mathsf{k_s}$ as follows $\mathsf{k_s}= \mathrm{KDF}(\mathsf{sk}\cdot C_0)$.  

\paragraph*{\textsc{SEnc} $((C_1, C_2)\leftarrow (\mathsf{k_s}, M))$} It takes the shared secret key $\mathsf{k_s}$ and a message $M$ as input. It outputs two ciphertext components $(C_1, C_2)$, where $C_1 = \textsc{Enc}(\mathsf{s_1}, M)$ is the encrypted component of the message $M$ using a symmetric key encryption algorithm, and $C_2 = \textsc{MAC}(\mathsf{s_2}, C_1)$ is a message authentication code to check the integrity and authenticity of $C_1$. Here, $(\mathsf{s_1}, \mathsf{s_2})$ are the leftmost and rightmost octets of the shared secret key $\mathsf{k_s}$.


\paragraph*{\textsc{SDec} $(M\leftarrow (\mathsf{k_s}, C_1, C_2))$} It takes the shared secret key $\mathsf{k_s}$, $C_1$, and $C_2$ as input. It outputs the actual message $M$ as follows: It extracts $(\mathsf{s_1}, \mathsf{s_2})$ from $\mathsf{k_s}$, verifies if $C_2 = \textsc{MAC}(\mathsf{s_2}, C_1)$, and if the verification is successful, decrypts $C_1$ to obtain $M$ using $\textsc{Dec}(\mathsf{s_1}, C_1)$.
\par 
Please refer to \cite{Shoup2001} for more details on the ECIES algorithms.

\section{Detailed Description of 5G-AKA Protocol} 
\label{appendix:5G-AKA}
We provide a detailed description of the 5G-AKA protocol here. We start with the \emph{Initiation} phase, followed by the \emph{Challenge-Response}, \emph{Sequence Number Re-synchronization}, and \emph{MAC Failure} phases. Figure \ref{fig:5G-AKA} shows a high-level overview of the 5G-AKA protocol.
\par
\textbf{Initiation}: This phase starts once the session between the subscriber and SN is initialized. The subscriber encrypts its unique identity $\mathrm{SUPI}$ with the HN's public key $\mathsf{PK_{HN}}$ using ECIES (please refer to Appendix \ref{sec:ECIES} for more details on ECIES) and produces a ciphertext $\mathrm{SUCI}$. The subscriber then sends $\mathrm{SUCI}$ and the identity of the HN, $\mathsf{ID_{HN}}$ to the SN. Afterward, the SN forwards the received $\mathrm{SUCI}$ to the HN, adding its own identity $\mathsf{ID_{SN}}$. Upon receiving $\mathrm{SUCI}$, the HN decrypts it using its own private key $\mathsf{sk_{HN}}$ and retrieves $\mathrm{SUPI}$, which helps the HN obtain the long-term secret key $\mathsf{k}$ and sequence number $\mathsf{SQN_{HN}}$ associated with the subscriber's $\mathrm{SUPI}$ from its database. 

\par 
\textbf{Challenge-Response}: In this phase, the subscriber and the HN mutually authenticate each other using a challenge-response method. Additionally, the subscriber and the SN establish the anchor keys, $\mathsf{K_{SEAF}}$, for any further secure communication.
\par 
The HN chooses a random challenge $R$ and generates a tuple $\left<R, \mathrm{AUTN}= \left<\mathrm{CONC}, \mathrm{MAC}\right>, \mathrm{HXRES}, \mathsf{K_{SEAF}}\right>$, which is then sent to the subscriber. In $\mathrm{CONC}$, the HN conceals the sequence number associated with the subscriber $\mathrm{SQN_{HN}}$ using the anonymous key $\mathsf{AK}$ derived from $R$ and $k$. The $\mathrm{MAC}$ is computed for the authentication and integrity checking of the random challenge $R$, using $R$, $k$, and $\mathrm{SQN_{HN}}$. Additionally, $\mathrm{HXRES}$ is computed by hashing $R$ with the expected response $\mathrm{XRES}$ from the subscriber, which is computed using $R$ and $k$. Finally, the anchor keys $\mathsf{K_{SEAF}}$ are computed using $k$, $R$, $\mathrm{ID_{SN}}$, and $\mathrm{SQN_{HN}}$. The HN also increments the sequence number $\mathrm{SQN_{HN}}$ by $1$ at the end.
\par 
After receiving the tuple $\left<R, \mathrm{AUTN}, \mathrm{HXRES}, \mathsf{K_{SEAF}}\right>$, the SN stores a copy of $\left<R, \mathrm{HXRES}, \mathsf{K_{SEAF}}\right>$ and forwards the tuple $\left<R, \mathrm{AUTN}\right>$ to the subscriber.
\par 
Upon receiving the tuple $\left<R, \mathrm{AUTN}\right>$, the subscriber computes an anonymous key $\mathsf{AK}$ inside the USIM card using $\mathsf{k}$ and $R$, and uncovers the sequence number $\mathrm{SQN_{HN}}$ from the $\mathrm{CONC}$ part of $\mathrm{AUTN}$. Next, it checks $\mathrm{MAC}$ using $R$ and $\mathrm{SQN_{HN}}$ inside the USIM card. If this check fails, it returns a \textsc{MAC\_Failure} message, which is sent to the SN and goes to the \emph{MAC Failure} phase. If the check is successful, the subscriber verifies the freshness of $\mathrm{SQN_{HN}}$. If this verification fails, the USIM card returns $\mathrm{AUTN}= \left<\mathrm{CONC, MAC^*}\right>$, where $\mathrm{CONC}$ conceals the sequence number $\mathrm{SQN_{UE}}$, and sends the tuple $\left<\textsc{Sync\_Failure}, \mathrm{AUTN}\right>$ to the HN to synchronize the sequence number. We shall present the re-synchronization process later. If the freshness check is successful, the USIM card sets $\mathrm{SQN_{UE}}$ to $\mathrm{SQN_{HN}}$, computes the anchor keys $\mathsf{K_{SEAF}}$ using $\mathsf{k}$, $R$, $\mathrm{ID_{SN}}$, and $\mathrm{SQN_{HN}}$, and finally returns the tuple $\left<\mathsf{K_{SEAF}}, \mathrm{RES}\right>$. The subscriber keeps $\mathsf{K_{SEAF}}$ and sends $\mathrm{RES}$ to the SN.
\par 
After receiving $\mathrm{RES}$, the SN compares the hash of $\mathrm{RES}$ and $R$ with $\mathrm{HXRES}$. If successful, the SN forwards $\mathrm{RES}$ to the HN, which then compares $\mathrm{RES}$ with its stored $\mathrm{XRES}$. If this comparison is successful and the subscriber is authenticated, the HN sends the $\mathrm{SUPI}$ associated with the subscriber to the SN. This concludes the 5G-AKA protocol execution for the current session. 
\par 
Please note that 3GPP TS 33.501 \cite{3GPP} also specifies that the subscriber and SN should implicitly confirm the agreed keys and each other's identities through the successful use of keys in subsequent procedures. This can be achieved with an additional key-confirmation round trip using $\mathsf{K_{SEAF}}$.
\par 
\textbf{Sequence Number Re-synchronization}: This phase is initiated when the subscriber needs to re-synchronize its sequence number with the HN. The primary purpose of using a sequence number-based freshness check is to prevent replay attacks. However, factors such as message loss or system failure may lead to a desynchronization between the subscriber's $\mathrm{SQN_{UE}}$ and the HN's $\mathrm{SQN_{HN}}$.
\par 
When the freshness check fails, as previously mentioned, the USIM card returns $\mathrm{AUTN}= \left<\mathrm{CONC, MAC^*}\right>$, with $\mathrm{CONC}$ concealing the sequence number $\mathrm{SQN_{UE}}$. It then sends the tuple $\left<\textsc{Sync\_Failure}, \mathrm{AUTN}\right>$ to the HN to synchronize the sequence number. Upon receiving the $\mathrm{Sync\_Failure}$ message, the SN forwards the tuple $\left<\textsc{Sync\_Failure}, \mathrm{AUTN}, R, \mathrm{SUCI}\right>$ to the HN. Subsequently, the HN de-conceals the $\mathrm{SQN_{UE}}$ after verifying the message authentication code $\mathrm{MAC^*}$. If the verification is successful, the HN sets its sequence number $\mathrm{SQN_{HN}}$ to $\mathrm{SQN_{UE}}+ 1$.

\par 
\textbf{MAC Failure}: This phase is initiated when the $\mathrm{MAC}$ check fails in the \emph{Challenge-Response} phase. In this phase, the subscriber simply returns a $\mathrm{MAC_Failure}$ message and goes to the \emph{Initiation} phase to restart the AKA protocol in a new session.

\section{Privacy Issues with 5G-AKA Protocol}
\label{appendix:5G-AKA-Privacy-Issue}
In this section, we briefly discuss the three main types of privacy-related attacks that can take place in the 5G-AKA protocol. Please note that these three types of attack are well explained in \cite{Wang2021}. The three types of privacy-related attacks the 5G-AKA protocol is vulnerable to are: \emph{Failure Message Linkability Attack} \cite{Arapinis2012}, \emph{Sequence Number Inference Attack} \cite{Borgaonkar2019}, and \emph{Encrypted SUPI Replay Attack} \cite{Fouque2016, Koutsos2019}. 
\par 
\emph{Failure Message Linkability Attack}: The goal of this attack is to distinguish a targeted subscriber from others by analyzing the responses received after replaying records of $\left<R, \mathrm{AUTN}\right>$ to all subscribers in the vicinity. The targeted subscriber responds with a $\textsc{Sync\_Failure}$ message because the replayed message passes the initial $\mathrm{MAC}$ verification with the correct long-term secret key $\mathsf{k}$ but fails the freshness check $\mathrm{SQN_{HN}}$. In contrast, other subscribers respond with a $\textsc{MAC\_Failure}$ message, as the replayed message fails the $\mathrm{MAC}$ verification due to a mismatched long-term secret key $\mathsf{k}$.
\par 

\emph{Sequence Number Inference Attack}: The objective of this attack is to learn information about the targeted subscriber's sequence number $\mathrm{SQN_{UE}}$. The attacker replays previously captured tuples $\left<R, \mathrm{AUTN}\right>$ multiple times and captures the returned $\mathrm{CONC}$ in the $\textsc{Sync\_Failure}$ messages. The attacker attempts to learn $\mathrm{SQN^i_{UE}} \oplus \mathrm{SQN^{i+1}_{UE}}$ by performing an Exclusive-OR operation between $\mathrm{CONC^i}$ and $\mathrm{CONC^{i+1}}$, as both $\mathrm{CONC^i}$ and $\mathrm{CONC^{i+1}}$ have concealed their sequence numbers using the same anonymous key $\mathsf{AK}$. 
\par 

\emph{Encrypted SUPI Replay Attack}: The goal of this attack is to distinguish the targeted subscriber from others. In this attack, the attacker replays the captured $\mathrm{SUCI}$ during the Initiation phase to the HN in all subscriber sessions and waits for the subscribers' responses to the corresponding challenge messages from the HN. The targeted subscriber responds successfully, while the other subscribers reply with $\textsc{MAC\_Failure}$ messages because the long-term secret key $\mathsf{k}$ used to generate the $\mathrm{MAC}$ matches only for the targeted subscriber and not for the others.

\end{document}